\documentclass[aps,prb,amsmath,amssymb,reprint,superscriptaddress]{revtex4-1}

\usepackage{hhline}
\usepackage{color}
\usepackage{graphicx}
\usepackage{subfigure}
\usepackage{verbatim}
\usepackage{fourier}
\usepackage{amsfonts}
\usepackage{multirow}
\usepackage{diagbox}
\usepackage{booktabs}

\begin{document}

\preprint{HIT-L2C-UM}

\title{Radiative heat transfer and radiative thermal energy for 2D nanoparticle ensembles}

\author{Minggang Luo}
\affiliation{School of Energy Science and Engineering, Harbin Institute of Technology, 92 West Street, Harbin 150001, China}
\affiliation{Laboratoire Charles Coulomb (L2C) UMR 5221 CNRS-Universit\'e de Montpellier, F- 34095 Montpellier, France}

\author{Junming Zhao}
\email[]{jmzhao@hit.edu.cn (Junming Zhao)}
\affiliation{School of Energy Science and Engineering, Harbin Institute of Technology, 92 West Street, Harbin 150001, China}
\affiliation{Key Laboratory of Aerospace Thermophysics, Ministry of Industry and Information Technology, Harbin 150001, China}

\author{Linhua Liu}
\affiliation{School of Energy and Power Engineering, Shandong University, Qingdao 266237, China}

\author{Mauro Antezza}
\email[]{mauro.antezza@umontpellier.fr (Mauro Antezza)}
\affiliation{Laboratoire Charles Coulomb (L2C) UMR 5221 CNRS-Universit\'e de Montpellier, F- 34095 Montpellier, France}
\affiliation{Institut Universitaire de France, 1 rue Descartes, F-75231 Paris Cedex 05, France}

\date{\today}

\begin{abstract}
Radiative heat transfer (RHT) and radiative thermal energy (RTE) for 2D nanoparticle ensembles are investigated in the framework of many-body radiative heat transfer theory. We consider nanoparticles made of different materials: metals (Ag), polar dielectrics (SiC) or insulator-metallic phase-change materials (VO$_2$). We start by investigating the RHT between two parallel 2D finite-size square-lattice nanoparticle ensembles, with particular attention to many-body interactions (MBI) effects. We fix the particles radius $a$ as the smallest length scale, and we describe the electromagnetic scattering from particles within the dipole approximation. 
Depending on the minimal distance between the in-plane particles (the lattice spacing $p$ for periodic systems), on the separation $d$ between the two lattice and on the thermal wavelength $\lambda_T=\hbar c/k_BT$, we systematically analyze the different  physical regimes characterizing the RHT. Four regimes are identified, \emph{rarefied regime},\emph{dense regime},\emph{non-MBI regime} and \emph{MBI regime}, respectively. When $p\ll \lambda_T^{}$, a multiple scattering of the electromagnetic field inside the systems gives rise to a MBI regime. MBI effects manifest themselves in different ways, depending  on the separation $d$: (a) if $d > \lambda_T$, due to the pure intra-ensemble MBI inside each 2D ensemble, the total heat conductance is less affected, and the thermal conductance spectrum manifests a single peak which is nonetheless shifted with respect to the one typical of two isolated nanoparticles. (b) if $d < \lambda_T$, there is a strong simultaneous intra- and inter-ensemble MBI. In this regime there is a direct quantitative effect on the heat conductance, in addition to a qualitative effect on the thermal conductance spectrum which now manifests a new second peak. As for the RTE, to correctly describe the radiation emitted by metallic nanoparticles, we derive an expression of the Poynting vector including also magnetic contribution, in addition to the electric one. By analyzing both periodic and non-periodic ensembles, we show that the RTE emitted by a single 2D nanoparticle ensemble is sensitive to the particle distribution. As instance, we see that the RTE emitted by 2D concentric ring-configuration ensemble has an inhibition feature near the center of the ensemble. 
\end{abstract}

\maketitle 

\section{Introduction}
Near-field radiative heat transfer (NFRHT) has recently attracted much attention for both fundamental and applicative reasons. When the separation distance between two objects is comparable to or less than the thermal wavelength $\lambda_T=\hbar c/k_BT$, near-field (evanescent waves) effect plays a dominant role in determining the net radiative heat exchange. The fluctuational electrodynamics theory developed by Rytov et al. \cite{Rytov1989} is the basic theoretical framework to analyze NFRHT.  NFRHT between two planar surfaces \cite{Polder1971,Loomis1994,Carminati1999,Shchegrov2000,Volokitin2001,Narayanaswamy2003,Volokitin2004}, two isolated nanoparticles \cite{Chapuis2008,Manjavacas2012,Nikbakht2018}, two spheres \cite{Narayanaswamy2008}, one dipole and surface \cite{Chapuis2008plate}, two nanoparticles above a substrate \cite{Messina2018,DongPrb2018,Zhang2019R,He2019APL} and between two nanoparticles separated by a multilayer plate \cite{Zhang2019T} were investigated theoretically recently in such framework. The theory has been set in a general framework, and it is now possible to calculate the NFRHT between two or many bodies of arbitrary shape and dielectric permittivity using a general scattering matrix method \cite{Messina2011,Messina2014}, which was then applied to the many-body system with planar geometry \cite{Latella2017prb}. On the experimental side, the radiative heat flux between two objects (e.g., two plates \cite{Ottens2011,Lim2015,Watjen2016,Ghashami2018,DeSutter2019}, one plate and one sphere or tip \cite{Shen2009,Rousseau2009,Song2015}) has been experimentally proven to be several orders of magnitude larger than the Planck\textquotesingle s black-body limit due to evanescent wave tunneling.

In this paper we consider the radiative heat transfer (RHT) between ensembles of nanoparticles. NFRHT in dense nanoparticle systems is difficult to determine due to the complex many-body interaction \cite{Ben2013,Tervo2017}. In a dense nanoparticle system, nanoparticles lie in the near field of each other, which leads to the significant multiple scattering of the thermally excited evanescent wave, hence the many-body interaction (MBI) will play a key role, and the two-body framework cannot be directly applied. To analyze NFRHT in systems composed of many nanoparticles, many theoretical frameworks have been developed, e.g., the many-body radiative heat transfer theory \cite{Ben2011,DongPrb2017}, scattering matrix method \cite{Messina2011,Messina2014}, trace formulas method \cite{Matthias2012,Muller2017}  and the quasi-analytic solution \cite{Czapla2019} based on the framework proposed to investigate NFRHT between two spheres \cite{Narayanaswamy2008}.

For NFRHT in the 3 dimensional (3D) nanoparticle system, some important progress has been reported recently. The MBI can not only enhance, but also inhibit, and even have nearly no effect on the radiative heat flux between dielectric many particle systems \cite{Ben2011,Dong2017JQ}, metallic many nanoparticle clusters \cite{Luo2019} and core-shell nanoparticle assemblies \cite{Chen2018JQ}, as well as the 1D nanoparticle chains \cite{Luo2019JQ}. By investigating NFRHT in the dense particle system from the point view of continuum medium, the heat super-diffusion was found in the plasmonic nanostructure networks due to NFRHT based on the fractional diffusion theory \cite{Ben2013}. Then, a similar heat super-diffusion was found in the periodically arranged planar SiC plates \cite{Latella2018}. Recently, a new method was developed to calculate the diffusive radiative thermal conductivity of arbitrary collections of nanoparticles \cite{Tervo2019}, which is an important progress relative to the kinetic method used to calculate the effective radiative thermal conductivity of 1D nanoparticle chain \cite{Ben2008,Tervo2016,Kathmann2018}. Also the radiative thermal energy (RTE) emitted in the near field  by a set of interacting nanoparticles has been the object of investigations, and has been recently predicted to focus the field in spots that are much smaller than those obtained with a single thermal source \cite{Ben2019}. 

For NFRHT in the 2 dimensional (2D) nanoparticle ensembles, some interesting phenomena have been reported. On the one hand, the inter-ensemble NFRHT between the 2D nanoparticle ensembles has been investigated in the extreme near field. The NFRHT between two gold nanoparticle array layers with 1 nm separation layer edge to layer edge was reported \cite{Phan2013}, where the multipole contribution to NFRHT should be considered. An interesting oscillatory-like feature of the NFRHT with translation of one array along its extending direction was observed. On the other hand, the investigation on intra-ensemble NFRHT in the 2D fractal nanoparticle ensembles also have been reported and the spatial distribution of nanoparticles in a 2D nanoparticle ensemble was demonstrated to play a key role in determining the radiative heat flux \cite{Nikbakht2017}. 

As for the RHT and radiative thermal energy for the 2D nanoparticle ensembles, there are several important aspects still needing investigation. Indeed, the way in which the MBI manifests itself in the total thermal conductance and in the thermal conductance spectrum, has not been investigated. A study of the combined effects coming from different 2D geometrical arrangements (periodic, non-periodic, concentric-rings) and dielectric properties (metals, polar dielectrics, metal-insulator phase-change materials ) of the nanoparticles ensembles is missing. We address these points in this paper, where the RHT  between 2D nanoparticle ensembles is investigated by means of many-body radiative heat transfer theory \cite{Ben2011}, further extended to metallic system in the coupled electric and magnetic dipole (CEMD) approach \cite{DongPrb2017}. We also study, for these systems, the RTE. 

This work is organized as follows. In Sec.\ref{Theorectical_aspect}, the CEMD approach is presented in brief, together with the expression for the RHT and thermal conductance. Concerning the RTE, the expression of the Poynting vector is derived for the general case where also magnetic dipoles are present, in addition to electric ones. This extension of the existing Poynting vector expression allows to describe the thermal emission from both dielectric and metallic nanoparticles. In addition, the physical model of the 2D nanoparticle ensemble and optical properties of the materials used in this work is also given. In Sec.\ref{asymptotic_regime_of_RHT}, asymptotic regimes of RHT between 2D finite-size square-lattice nanoparticle ensembles are summarized, as well as the simplified formulas for the thermal conductance in different regimes. In Sec.\ref{RHF_2D_ensemble}, RHT between 2D periodic nanoparticle ensembles are analyzed, with particular attention devoted to MBI effects. The effects of metal-insulator phase change of the nanoparticles  and of the lateral translation of the two parallel 2D ensembles are also analyzed.  In Sec.~\ref{RTE_2D_ensemble}, RTE emitted by the single 2D nanoparticle ensemble is analyzed, as a function of the particle distribution (e.g., periodic, random and concentric ring configurations) and dielectric properties. A study of the relative weight of the magnetic dipole and electric dipole contributions to RTE is also conducted. 
 
\section{Theoretical models}

\label{Theorectical_aspect}
In this section, we describe the physical systems, the theoretical models for the RHT and RTE, and finally the optical properties of the materials we use.

\subsection{Physical systems: 2D nanoparticle ensembles}
In this paper we will investigate RHT between 2D nanoparticle ensembles by studying radiative thermal conductance between two parallel 2D finite-size square-lattice nanoparticle ensembles (Secs.\ref{RHF_2D_ensemble}) and RTE emitted by a single 2D periodic and non-periodic nanoparticle ensemble (Sec.~\ref{RTE_2D_ensemble}). We will hence introduce two physical systems, respectively.

\begin{figure} [htbp]
\centerline {\includegraphics[width=0.45\textwidth]{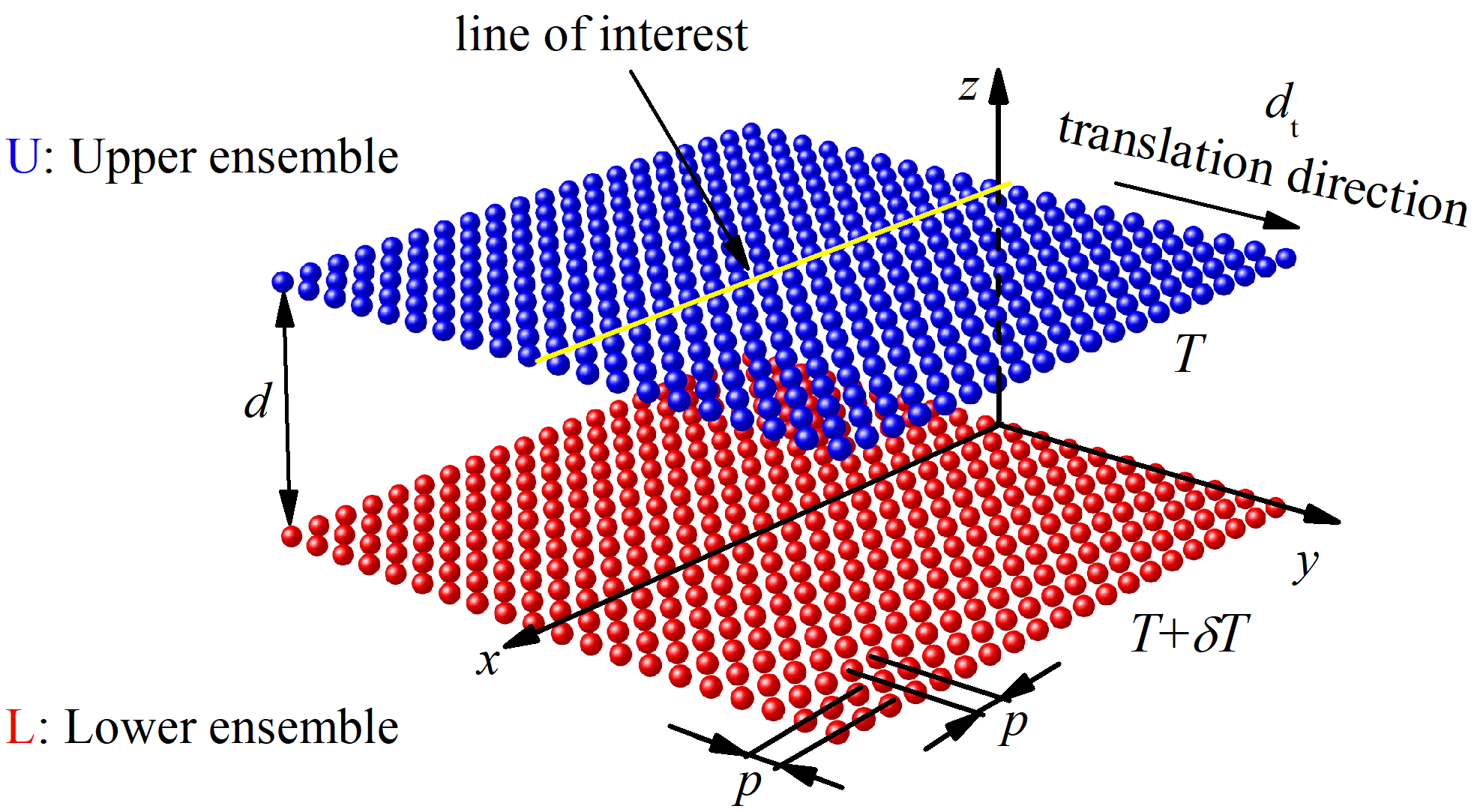}}
\caption{Schematic of the 2D finite-size square-lattice nanoparticle ensembles. RHT between the upper nanoparticle ensemble (U, at temperature $T$) and the lower nanoparticle ensemble (L, at temperature $T+\delta T$) is investigated. Nanoparticle radius is $a$. The separation distance between the ensemble L and U center to center is $d$. $d_{\textrm{t}}^{}$ is the translation distance of the ensemble U relative to the ensemble L. Periodicity of the periodically distributed nanoparticle ensemble is $p$. $N$ is the number of nanoparticles in each ensemble.}
\label{Structure_diagram}
\end{figure}

Concerning the RHT, we will focus on parallel 2D finite-size square-lattice nanoparticle ensembles, as shown in Fig.~\ref{Structure_diagram}.

The radiative thermal conductance between the lower ensemble (L, at temperature $T+\delta T$) and upper ensemble (U, at temperature $T$) is calculated with various separation distance $d$ center to center. In general, each ensemble is composed of $N=400$ nanoparticles. 
The nanoparticle radius is $a$. When needed, we will evaluate the energy density and Poynting vector along the yellow line shown in the Fig.~\ref{Structure_diagram}. When investigating the effect of translation of particle ensemble on RHT, the lower ensemble is fixed and the upper ensemble is translated along the translation direction. The origin of the Cartesian coordinate system is fixed at the center of an edge of the lower ensemble and the particle ensemble is periodically distributed $~20 \times 20$ nanoparticles ensemble. The periodicity $p$ is the separation distance between two neighboring particles center to center in the line parallel to the edge of the ensemble.

\begin{figure} [htbp]
     \centering
     \subfigure [Periodically distributed particle ensemble] {\includegraphics[width=0.225\textwidth]{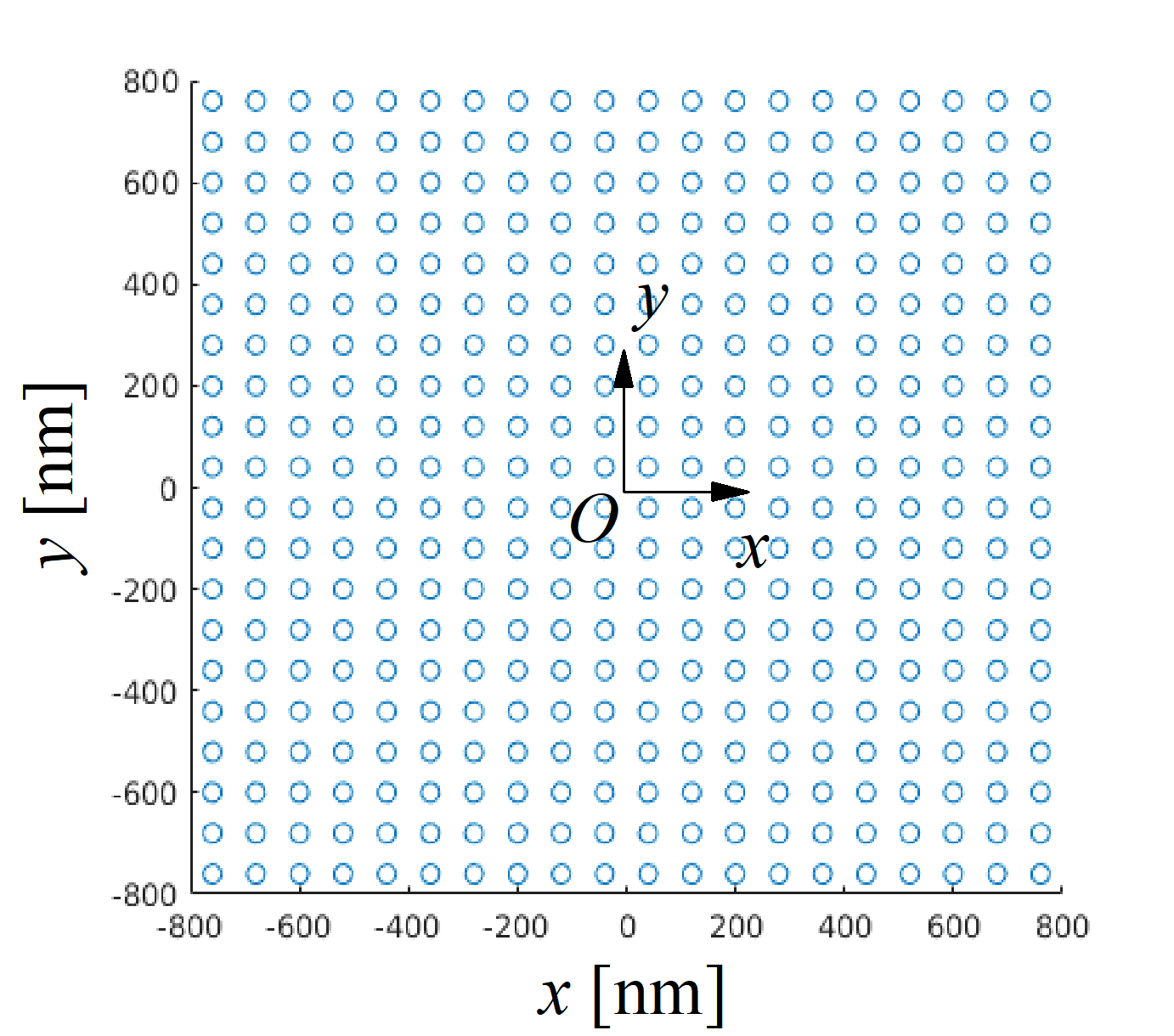}}
     \hspace{8pt}
     \subfigure [Randomly distributed particle ensemble] {\includegraphics[width=0.232\textwidth]{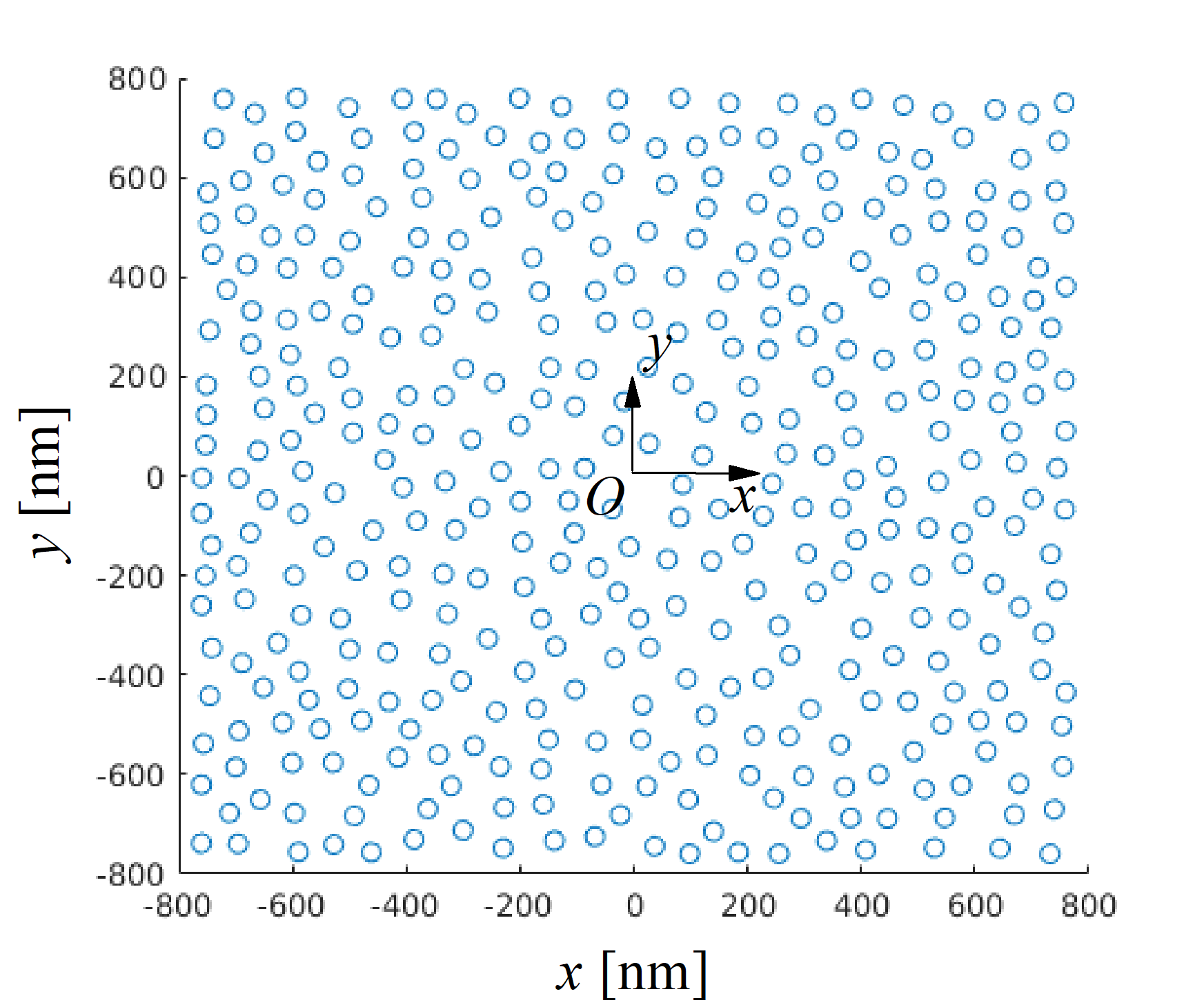}}
          \hspace{8pt}
     \subfigure [Concentric ring configuration particle ensemble] {\includegraphics[width=0.28\textwidth]{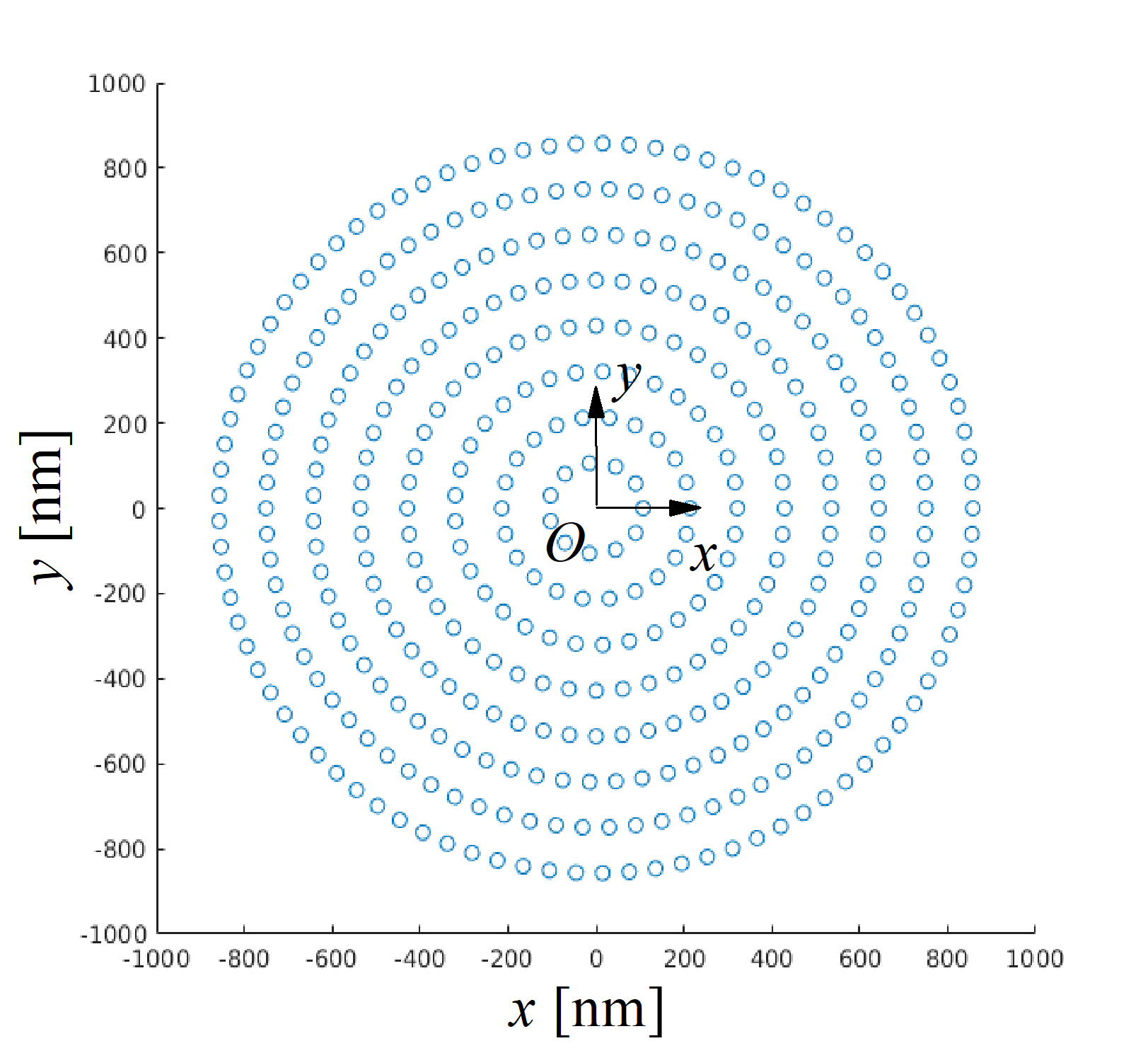}}
        \caption{Three kinds of nanoparticle distribution: (a) periodically distributed particle ensemble; (b) randomly distributed particle ensemble and (c) concentric ring configuration particle ensemble. $N$ = 400, $a=20$ nm.}
        \label{2dstructure}
\end{figure}

When we discuss the RTE emitted by the single nanoparticle ensemble (Sec.~\ref{RTE_2D_ensemble}), we will refer to  the physical systems of Fig.~\ref{2dstructure}, where three different kinds of 2D nanoparticle distributions are considered: (a) periodic ensemble; (b) random ensemble and (c) concentric ring configuration. The origin of the Cartesian coordinate system is set at the center of the ensemble. We will consider $N = 400$ nanoparticles of radius $a=20$ nm.

\subsection{Radiative heat transfer and thermal conductance between nanoparticle ensembles}
Let us start considering two particles. In the framework of the CEMD approach, the power absorbed by the $i$th particle and radiated by $j$th particle is written in the Landauer formalism \cite{Luo2019,DongPrb2017,Ben2011}
\begin{equation}
\varphi_{j\rightarrow i}^{}=3\int_{0}^{+\infty} \frac{\mathrm{d}\omega}{2\pi}\Theta(\omega,T_j)\mathcal{T}_{i,j}(\omega),
\label{power}
\end{equation}
where $\omega$ is electromagnetic field angular frequency, $\Theta(\omega,T_j)$ is the mean energy of the harmonic Planck\textquotesingle s oscillator and $\mathcal{T}_{i,j}(\omega)$ is transmission coefficient between the $j$th and $i$th particles, given by \cite{Luo2019,DongPrb2017}:
\begin{equation}
\begin{aligned}
\mathcal{T}_{i,j}(\omega)=&\frac{4}{3}k^4\left[\textrm{Im}(\chi_E^i)\textrm{Im}(\chi_E^j)\textrm{Tr}(G_{ij}^{EE}G_{ij}^{EE\dagger})\right.\\&+\textrm{Im}(\chi_E^i)\textrm{Im}(\chi_H^j)\textrm{Tr}(G_{ij}^{EM}G_{ij}^{EM\dagger})\\&+\textrm{Im}(\chi_H^i)\textrm{Im}(\chi_E^j)\textrm{Tr}(G_{ij}^{ME}G_{ij}^{ME\dagger})\\&+\left.\textrm{Im}(\chi_H^i)\textrm{Im}(\chi_H^j)\textrm{Tr}(G_{ij}^{MM}G_{ij}^{MM\dagger})\right].
\label{transmission}
 \end{aligned}
\end{equation}
Here parameters $\chi_E^{}=\alpha_E^{}-\frac{ik^3}{6\pi}
\left|\alpha_E^{}\right|^2$ and $\chi_H^{}=\alpha_H^{}-\frac{ik^3}{6\pi}
\left|\alpha_H^{}\right|^2$ are introduced \cite{Manjavacas2012}, $\alpha_E$ and $\alpha_H$ are electric and magnetic dipole polarizabilities, $k$ is the wavevector in vacuum, $G_{ij}^{\nu\tau}$ ($\nu$,$\tau$=$E$~or~$M$) is the Green\textquotesingle s function for the many-particle system considering the many-body interaction, which is the solution of Eq.~(\ref{double_sources}) in the Appendix. The net power exchanged between these two nanoparticles (radiative heat flux) is 
\begin{equation}
\begin{aligned}
\varphi_{j\leftrightarrow i}^{}&=\varphi_{j\rightarrow i}^{}-\varphi_{i\rightarrow j}^{}\\
&=3\int_{0}^{+\infty} \frac{\mathrm{d}\omega}{2\pi}\left(\Theta(\omega,T_j)-\Theta(\omega,T_i)\right)\mathcal{T}_{i,j}(\omega),
\label{ExchangedPower}
\end{aligned}
\end{equation}
which allows to define the thermal conductance ($G$) between the particle $i$ and $j$ as \cite{Dong2017JQ,Luo2019}
\begin{equation}
G_{ij}^{}=\lim_{\delta T \rightarrow 0}\frac{\varphi_{j\leftrightarrow i}^{}}{T_j-T_i} ,
\label{Gij}
\end{equation}
where $\delta T=T_j-T_i$ is the temperature difference between the two particles. Let us now consider finite-size square-lattice nanoparticle ensembles, L and U, as shown in Fig.~\ref{Structure_diagram}. We can define the radiative heat flux between the two particle ensembles as
\begin{equation}
\begin{aligned}
\varphi&=\sum_{i\in \textrm{U}}^{}\sum_{j\in \textrm{L}}^{}\varphi_{j\leftrightarrow i}^{} ,
\label{RHF}
\end{aligned}
\end{equation}
where nanoparticle $i$ and $j$ belong to the upper nanoparticle ensemble (U) and the lower nanoparticle ensemble (L), respectively. The total thermal conductance ($G$) between the ensemble U and ensemble L with a separation is a function of many parameters (e.g., temperature $T$, separation $d$ and lattice spacing $p$, etc.), which is defined as follows. 
\begin{equation}
G(p,d)=\lim_{\delta T \rightarrow 0}\frac{\varphi}{\delta T}=\sum_{i\in \textrm{U}}^{}\sum_{j\in \textrm{L}}^{}G_{ij}^{}.
\label{Gt}
\end{equation}
The total thermal conductance $G(p,d)$ between the two ensembles is the sum of the thermal conductances $G_{ij}$ of all possible nanoparticle pairs, where one nanoparticle is from the ensemble L and the other one is from the ensemble U. The total thermal conductance can also be expressed as the frequency integral of the spectral thermal conductance $G_{\omega}(p,d)$: $G(p,d)=\int_{0}^{+\infty} G_{\omega}(p,d)~\text{d}\omega$.

\subsection{Radiative thermal energy emitted by the single nanoparticle ensemble}
The radiative thermal energy at position $\mathbf{r}$ emitted by the particle ensemble can be evaluated by the Poynting vector, which is defined as follows \cite{Francoeur2008FDT,Ben2019}.
\begin{equation}
\left \langle \mathbf{S}(\mathbf{r}) \right \rangle=2\int_{0}^{+\infty} \left \langle \mathbf{S}(\mathbf{r},\omega) \right \rangle\, \frac{\text{d}\omega}{2\pi} ,
\label{S}
\end{equation}
where the spectral Poynting vector $\left \langle \mathbf{S}(\mathbf{r},\omega) \right \rangle$ yields
\begin{equation}
\left \langle \mathbf{S}(\mathbf{r},\omega) \right \rangle=\Re \left \langle \mathbf{E}(\mathbf{r},\omega) \times \mathbf{H}^*(\mathbf{r},\omega) \right \rangle ,
\label{S_spectral}
\end{equation}
where electric and magnetic field radiated by the fluctuating electric dipole ($\mathbf{p}^f$) and magnetic dipole ($\mathbf{m}^f$) yield
\begin{equation}
\mathbf{E}(\mathbf{r},\omega)=\mu_0\omega^2\sum_{i=1}^{N}G^{EE}(\mathbf{r} ,\mathbf{r}_i)\mathbf{p}_i^f+\mu_0\omega k\sum_{i=1}^{N}G^{EM}(\mathbf{r},\mathbf{r}_i)\mathbf{m}_i^f ,
\label{dipole_E}
\end{equation}
\begin{equation}
\mathbf{H}(\mathbf{r},\omega)=k\omega\sum_{i=1}^{N}G^{ME}(\mathbf{r},\mathbf{r}_i)\mathbf{p}_i^f+k^2\sum_{i=1}^{N}G^{MM}(\mathbf{r},\mathbf{r}_i)\mathbf{m}_i^f ,
\label{dipole_H}
\end{equation}
where $\mu_0$ is the vacuum permeability, $N$ is the number of particles. $G^{\nu\tau}(\mathbf{r},\mathbf{r}_i)$ ($\nu$,$\tau$=$E$~or~$M$) is the Green\textquotesingle s function connecting the field point $\textbf{r}$ and the source point $\textbf{r}_i^{}$ in the particle system considering MBI, which is the solution of the Eq.~(\ref{single_source}) with the help of the Eqs.~(\ref{Compact_system_green_function_single_source}) and~(\ref{Compact_free_space_green_function_single_source}) in the Appendix. Substituting the Eqs.(\ref{dipole_E}) and (\ref{dipole_H}) into the Eq.~(\ref{S_spectral}), the spectral Poynting vector can be rearranged as
\begin{widetext}
\begin{equation}
\begin{aligned}
\left \langle \mathbf{S}(\mathbf{r},\omega) \right \rangle=&\Re \left \langle \mathbf{E}(\mathbf{r},\omega) \times \mathbf{H}^*(\mathbf{r},\omega) \right \rangle=\Re \left\{\left \langle \mathbf{x}\left(E_yH_{z}^*-E_zH_{y}^*\right)+\mathbf{y}\left(E_zH_{x}^*-E_xH_{z}^*\right)+\mathbf{z}\left(E_xH_{y}^*-E_yH_{x}^*\right) \right \rangle\right\}\\=&\sum_{i=1}^{N}\sum_{n=1}^{3}\sum_{m=1}^{3}\Re \left\{\mu_0\omega^3k\left[\mathbf{x}\left(G_{yn}^{EE}G_{zm}^{ME*}-G_{zn}^{EE}G_{ym}^{ME*}\right) +\mathbf{y}\left(G_{zn}^{EE}G_{xm}^{ME*}-G_{xn}^{EE}G_{zm}^{ME*}\right)+\mathbf{z}\left(G_{xn}^{EE}G_{ym}^{ME*}-G_{yn}^{EE}G_{xm}^{ME*}\right)\right]
\left \langle p_{i,n}^{f}p_{i,m}^{f*}\right \rangle\right.\\&+\left.\mu_0\omega k^3\left[\mathbf{x}\left(G_{yn}^{EM}G_{zm}^{MM*}-G_{zn}^{EM}G_{ym}^{MM*}\right)+\mathbf{y}\left(G_{zn}^{EM}G_{xm}^{MM*}-G_{xn}^{EM}G_{zm}^{MM*}\right)+\mathbf{z}\left(G_{xn}^{EM}G_{ym}^{MM*}-G_{yn}^{EM}G_{xm}^{MM*}\right)\right]
\left \langle m_{i,n}^{f}m_{i,m}^{f*}\right \rangle\right\} ,
\label{S_spectral_detail}
\end{aligned}
\end{equation}
\end{widetext}
where subscripts $m$ and $n$ are polarization direction index, $\mathbf{x}$, $\mathbf{y}$ and $\mathbf{z}$ are the unit vectors of $x$, $y$ and $z$ axes in the given Cartesian coordinate system. $G_{\mu_1 \mu_2}^{\nu \tau}$ ($\mu_1 =x, y, z$; $\mu_2=(m~\textrm{or}~n)=1,2,3$ and $\nu,\tau=E,M$) is the element of the $3~\times~3$ Green\textquotesingle s function $G^{\nu\tau}(\mathbf{r},\mathbf{r}_i)$ ($\nu$,$\tau$=$E$~or~$M$), which is the solution of the Eq.~(\ref{single_source}) with the help of the Eqs.~(\ref{Compact_system_green_function_single_source}) and~(\ref{Compact_free_space_green_function_single_source}) in the Appendix. The fluctuation dissipation theorem yields \cite{Ben2011,DongPrb2017,Luo2019}
\begin{equation}
\left \langle p_{i,n}^{f}p_{i,m}^{f*}\right \rangle=2\frac{\epsilon_0}{\omega}\Im\left(\chi_E^{}\right)\Theta(\omega,T)\delta_{nm} ,
\label{FDT_E}
\end{equation}
\begin{equation}
\left \langle m_{i,n}^{f}m_{i,m}^{f*}\right \rangle=\frac{2}{\mu_0\omega}\Im\left(\chi_H^{}\right)\Theta(\omega,T)\delta_{nm}.
\label{FDT_M}
\end{equation}
Finally, the spectral Poynting vector of Eq.~(\ref{S_spectral_detail}) can be rewritten as follows.
\begin{equation}
\left \langle \mathbf{S}(\mathbf{r},\omega) \right \rangle=2\sum_{i=1}^{N}\Re \left(k^3~\mathbb{S}~\Theta(\omega,T(\mathbf{r}_i))\right) ,
\label{S_short}
\end{equation}
where $\mathbb{S}$ is defined as follows:
\begin{widetext}
\begin{equation}
\begin{aligned}
\mathbb{S}=&\sum_{n=m=1}^{3} \left[\mathbf{x}\left(G_{yn}^{EE}G_{zm}^{ME*}-G_{zn}^{EE}G_{ym}^{ME*}\right)+\mathbf{y}\left(G_{zn}^{EE}G_{xm}^{ME*}-G_{xn}^{EE}G_{zm}^{ME*}\right)+\mathbf{z}\left(G_{xn}^{EE}G_{ym}^{ME*}-G_{yn}^{EE}G_{xm}^{ME*}\right)\right] \Im\left(\chi_E^{}\right)\\+&\sum_{n=m=1}^{3}\left[\mathbf{x}\left(G_{yn}^{EM}G_{zm}^{MM*}-G_{zn}^{EM}G_{ym}^{MM*}\right)+\mathbf{y}\left(G_{zn}^{EM}G_{xm}^{MM*}-G_{xn}^{EM}G_{zm}^{MM*}\right)+\mathbf{z}\left(G_{xn}^{EM}G_{ym}^{MM*}-G_{yn}^{EM}G_{xm}^{MM*}\right)\right]\Im\left(\chi_H^{}\right).
\label{S_spectral_final}
\end{aligned}
\end{equation}
\end{widetext}
Poynting vector emitted by the fluctuating electric and magnetic dipoles can be obtained by the Eqs.~(\ref{S_short}) and (\ref{S_spectral_final}), which is an extension of the recent work ~\cite{Ben2019} to take the magnetic dipole contribution into consideration and is applicable for not only dielectric but also metallic nanoparticle ensembles.

\subsection{Dielectric function and polarizability of nanoparticle}
Three different materials are used in present work, metallic Ag, dielectric SiC and phase-change VO$_2$, respectively. The dielectric functions of Ag and SiC are described by the Drude model $\epsilon(\omega) = 1-\omega_p^2/(\omega^2+i\gamma\omega)$ with parameters $\omega_p = 1.37 \times 10^{16}$ rad$\cdot$s$^{-1}$ and $\gamma$ = 2.732 $\times$ 10$^{13}$ rad$\cdot$s$^{-1}$ \cite{Ordal} and the Drude-Lorentz model $\epsilon(\omega) =\epsilon_{\infty}^{}(\omega^2-\omega_l^2+i\gamma\omega)/(\omega^2-\omega_t^2+i\gamma\omega)$ with parameters $\epsilon_{\infty}^{}$ = 6.7, $\omega_l^{}$ = 1.827 $\times$ 10$^{14}$ rad$\cdot$s$^{-1}$, $\omega_t^{}$ = 1.495 $\times$ 10$^{14}$ rad$\cdot$s$^{-1}$, and $\gamma$ = 0.9 $\times$ 10$^{12}$ rad$\cdot$s$^{-1}$ \cite{Palik}, respectively. The VO$_2$ is a kind of phase-change material, which undergoes an insulator-metal transition around 341 K (phase transition temperature). Below 341 K, VO$_2$ is an uniaxial anisotropic insulator, of which the dielectric function can be described by a tensor as follows
\begin{equation}
\begin{pmatrix}
\epsilon_\rVert^{} & 0 & 0\\
0 & \epsilon_\bot^{} & 0\\
0 & 0 & \epsilon_\bot^{}
\end{pmatrix},
\label{dielectric_tensor}
\end{equation}
where $\epsilon_\bot^{}$ and $\epsilon_\rVert^{}$ are ordinary and extraordinary dielectric function component relative to the optic axis of uniaxial insulating VO$_2$, respectively. Both ordinary and extraordinary dielectric function can be described by the Lorentz model as follows.
\begin{equation}
\epsilon(\omega)=\epsilon_{\infty}^{}+\sum_{k=1}^{N_L}\frac{S_k\omega_k^2}{\omega_k^2-i\gamma_k^{}\omega-\omega^2},
\label{Lorentz_model_VO}
\end{equation}
where $S_k$, $\omega_k^{}$ and $\gamma_k^{}$ are phonon strength,  phonon frequency and damping coefficient of the $k^{th}$ phonon mode. $N_L$ is the number of phonon modes. All the necessary parameters for both $\epsilon_\bot^{}$ and $\epsilon_\rVert^{}$ can be found in the Ref.\cite{VO21966}. Above 341K, VO$_2$ is an isotropic metal, of which the dielectric function can be described by a Drude model as follows \cite{VO21966}.
\begin{equation}
\epsilon(\omega)=\epsilon_{\infty}^{}\frac{\omega_p^2}{\omega^2-i\omega\gamma},
\label{Drude_model_VO}
\end{equation}
where $\epsilon_{\infty} = 9$, $\omega_p$ = $1.51\times10^{15}$ rad$\cdot$s$^{-1}$ and $\gamma$ =$1.88\times10^{15}$ rad$\cdot$s$^{-1}$.

For a nanoparticle composed of an isotropic material (e.g., metallic Ag and dielectric SiC), the electric and magnetic dipole polarizabilities are given as follows \cite{Chapuis2008}.
\begin{equation}
\alpha_E^{}=4\pi a^{3}\frac{\varepsilon-1}{\varepsilon+2},
\label{CM_E}
\end{equation}
where $\varepsilon$ is the relative permittivity and
\begin{equation}
\alpha_H^{}=\frac{2\pi}{15}a^{3}\left(\frac{\omega a}{c}\right)^2\left(\epsilon-1\right).
\label{CM_H}
\end{equation}
The polarizability of Ag and SiC nanoparticle can be found in our previous work \cite{DongPrb2017,Luo2019,Luo2019JQ}.

\begin{figure} [htbp]
     \centering
     \subfigure [Insulator VO$_2$] {\includegraphics[scale=0.35]{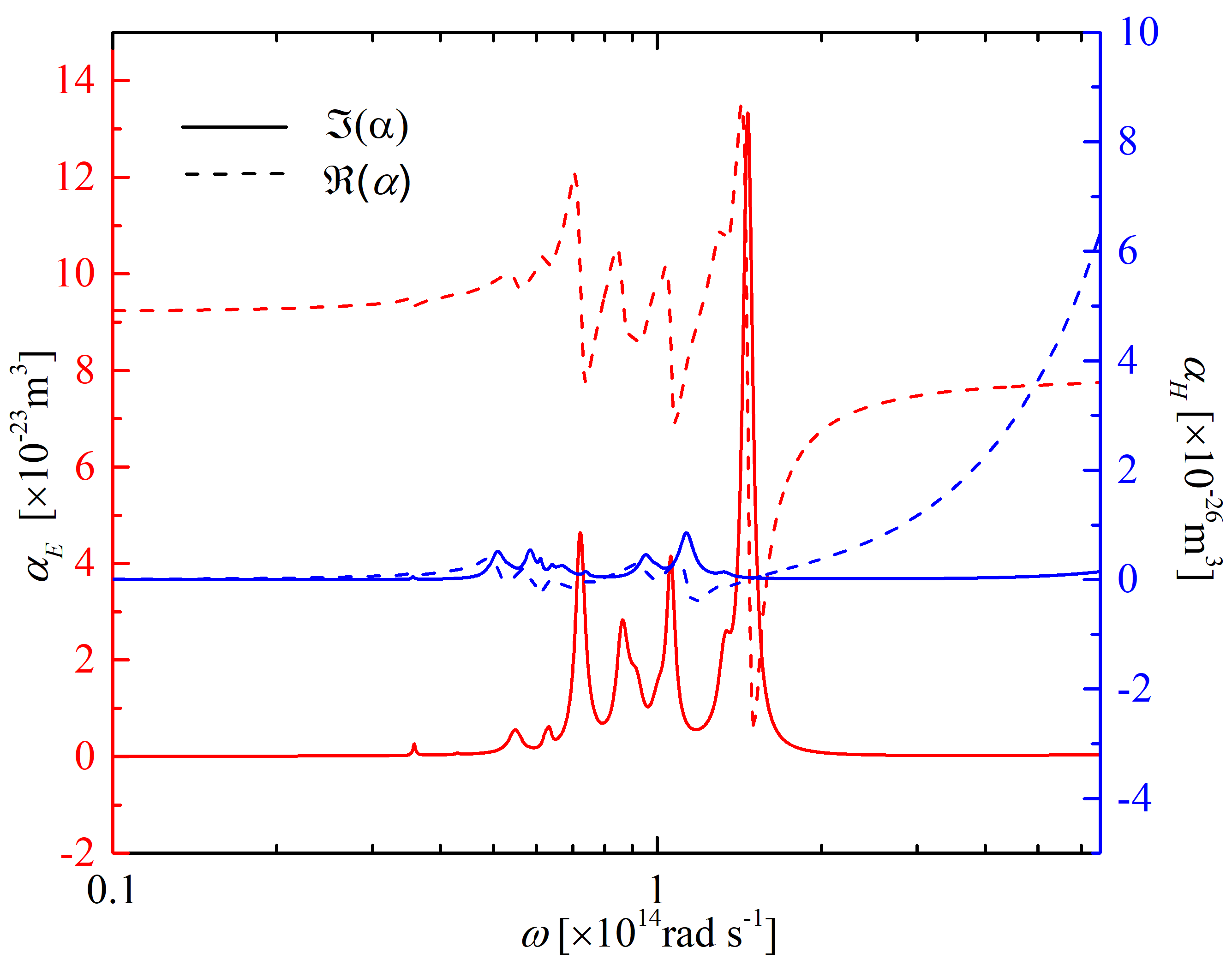}}\\
     \hspace{8pt}
     \subfigure [metallic VO$_2$] {\includegraphics[scale=0.35]{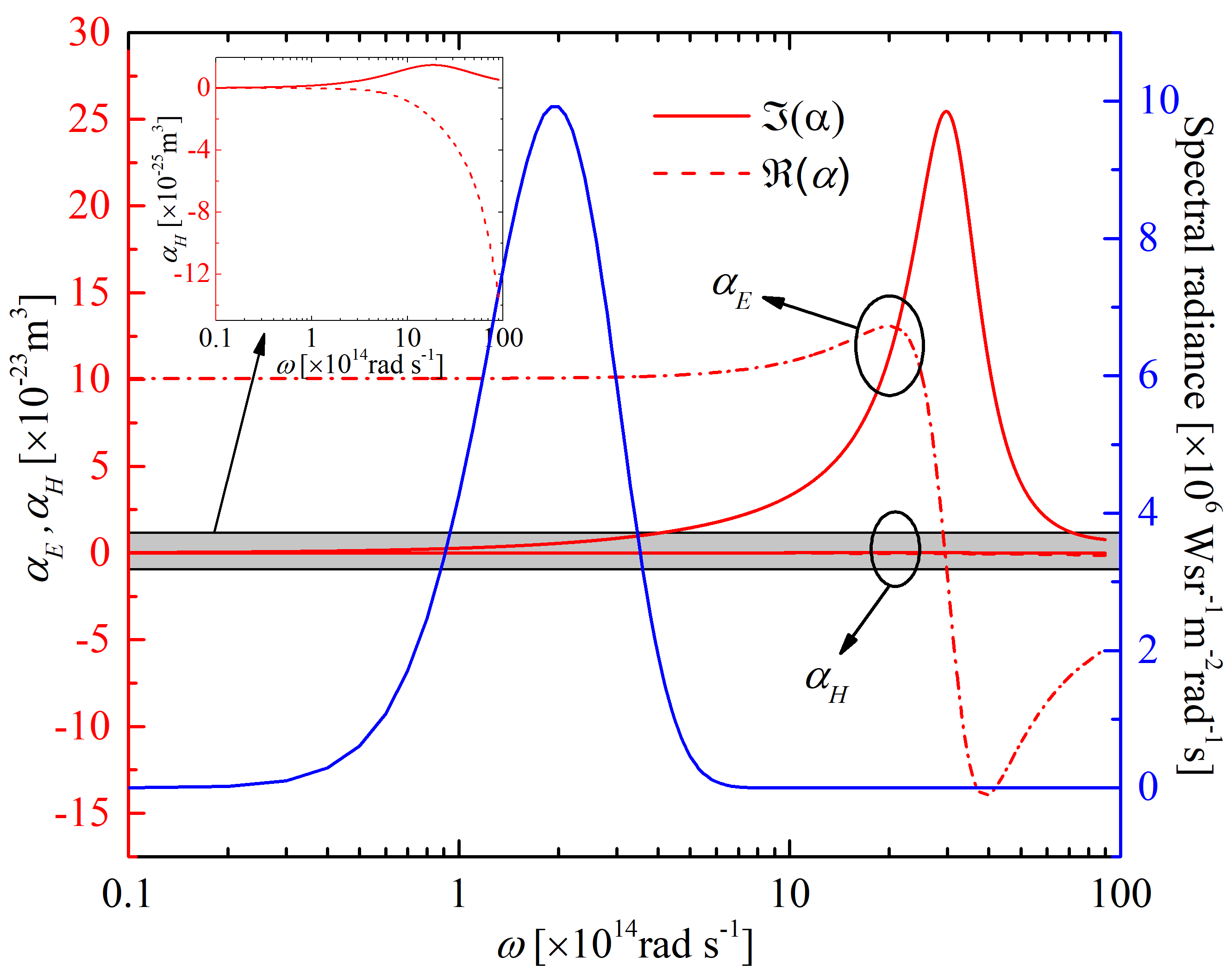}}
        \caption{The electric and magnetic polarizabilities for both the (a) insulator-phase VO$_2$ nanoparticle and (b) metallic-phase VO$_2$ nanoparticle. Nanoparticle radius $a$ is 20 nm. For insulator VO$_2$ particle, the ``$1\setminus 3-2\setminus 3$'' rule is applied to calculate the polarizability with the help of $\epsilon_\rVert^{}$ and $\epsilon_\bot^{}$ \cite{Quinten2010}. The spectral radiance of the blackbody at room temperature is also added for reference.}
        \label{polarizability}
\end{figure}

However, for the anisotropic insulator-phase VO$_2$ nanoparticle below the transition temperature, a well-established solution is applied to the polarizability of anisotropic spherical nanoparticle in two steps: first calculate polarizability for nanoparticle using $\epsilon_\bot^{}$ and $\epsilon_\rVert^{}$ separately, and then add up the results according to the $1\setminus 3-2\setminus 3$ rule \cite{Quinten2010}:
\begin{equation}
\alpha_{\nu}^{}=\frac{2}{3}\alpha_{\nu}^{}(\epsilon_\bot^{})+\frac{1}{3}\alpha_{\nu}^{}(\epsilon_\rVert^{}),
\label{CM_ave}
\end{equation}
where $\nu= E~\textrm{or}~H$ . The electric and magnetic polarizabilities for both insulator-phase and metallic-phase VO$_2$ nanoparticles are shown in Fig.~\ref{polarizability}. In order to compare the resonance frequency to the characteristic thermal frequency, the spectral radiance of the blackbody at room temperature is also added in Fig.~\ref{polarizability}(b) for reference. There is a mismatch between the characteristic thermal frequency and the polarizability resonance frequency of metallic VO$_2$ nanoparticle. A similar mismatch between characteristic thermal frequency and polarizability resonance frequency of Ag nanoparticle as that of metallic VO$_2$ nanoparticle can also be observed \cite{Luo2019JQ}. 

\section{Radiative heat transfer between 2D finite-size periodic square-lattice nanoparticle ensembles: \\ Asymptotic regimes}
\label{asymptotic_regime_of_RHT}
In this section we discuss the main asymptotic regimes of the RHT between 2D periodic finite-size square-lattice nanoparticle ensembles. In general, we have four length scales in this problem. Three of them are \emph{geometric} length scales: the lattice spacing $p$, the separation $d$ between the 2D systems, and the nanoparticle radius $a$. The fourth is a \emph{thermal} length scale: the characteristic thermal wavelength $\lambda_T^{}$. In this work, we keep fixed the value of $a=20$ nm, and set all the 3 other length scales much larger than $a$. We will see (section \ref{periodicity_effect_on_RHT}) that a fifth additional length scale, related to emergence of multiple scattering of the electromagnetic between the two nanoparticles of a pair,  will naturally emerge in this problem.

\begin{figure} [htbp]
\centerline {\includegraphics[scale=0.35]{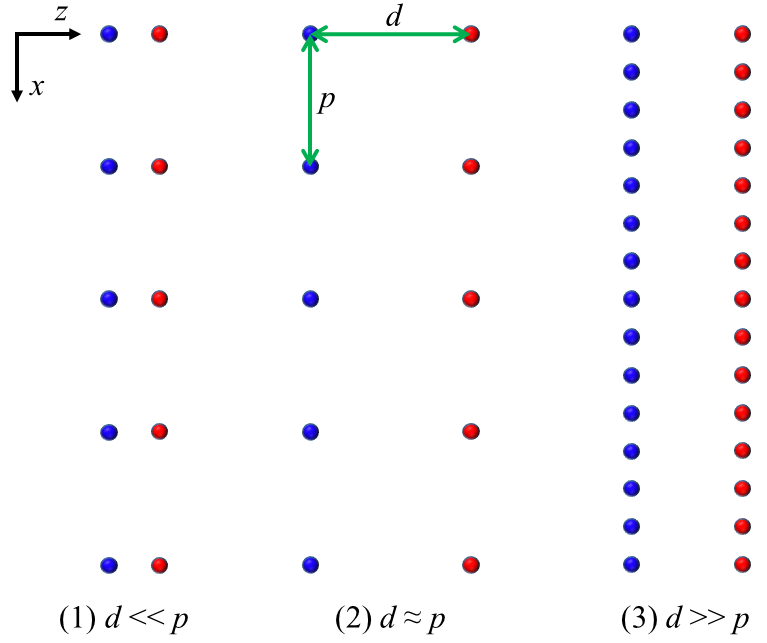}}
\caption{Depending on the relative values of $\{p,d\}$, 3 kinds of 2D finite-size square-lattice nanoparticle ensembles are possible: (1) $d\ll p$, (2) $d\approx p$, and (3) $d\gg p$.}
\label{regime_p_d}
\end{figure}

According to combinations of the geometric length scales $\{p,d\}$, 3 kinds of 2D finite-size square-lattice nanoparticle ensembles can be clarified, which are shown in Fig.~\ref{regime_p_d}.

\begin{figure} [htbp]
\centerline {\includegraphics[scale=0.45]{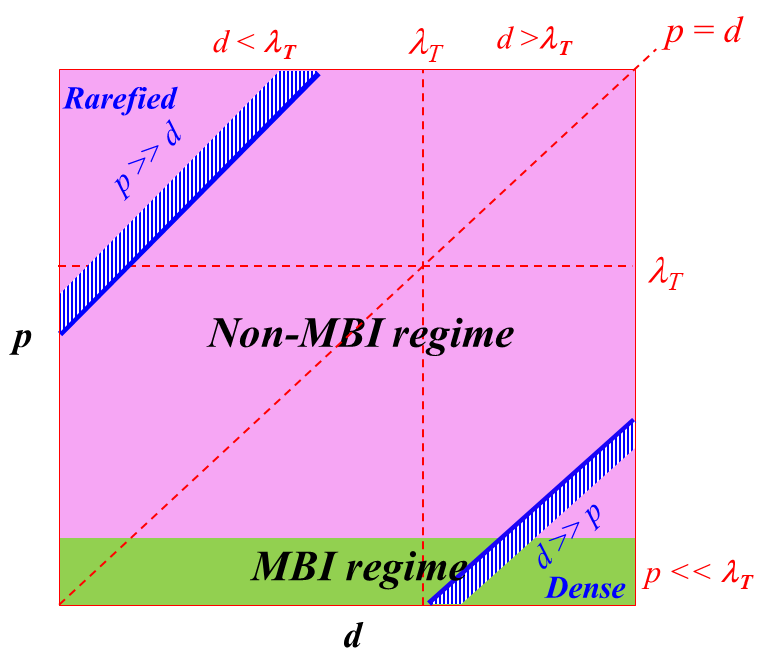}}
\caption{Asymptotic regime map of RHT between 2D finite-size square-lattice nanoparticle ensembles. Combined with the Table~\ref{MBI_NFE_regime}, four asymptotic regimes are identified, \emph{MBI regime}, \emph{non-MBI regime}, \emph{rarefied regime} and \emph{dense regime}, respectively.}
\label{regime_map_RHT}
\end{figure}

The MBI effects and near-field effects (NFE) on RHT are significantly dependent on the combination of $\{p, d, \lambda_T^{}\}$, and different asymptotic regimes can be identified, as listed in the Table~\ref{MBI_NFE_regime} and qualitatively shown in the regime map Fig.~\ref{regime_map_RHT}. Four asymptotic regimes are introduced, \emph{MBI regime}, \emph{non-MBI regime}, \emph{rarefied regime} and \emph{dense regime}, respectively. The terms \emph{intra-ensemble} and \emph{inter-ensemble} mean MBI effect and near-field effect inside the single nanoparticle ensemble and between the two nanoparticle ensembles, respectively. In the table, "yes" and "no" mean strong and negligible MBI effect and NFE, respectively.

\begin{table*}[htbp]
  \caption{Asymptotic regimes of RHT between 2D finite-size square-lattice nanoparticle ensembles}
  \centering
  \begin{tabular}{|c|c!{\vrule width 2pt}c|c|c|c|c|c|c|c|}
   \hline
     \multicolumn{2}{|c!{\vrule width 2pt}}{{Geometric conditions}}& \multicolumn{3}{c|}{$d\ll p$}& \multicolumn{2}{c|}{$d\approx p$}& \multicolumn{3}{c|}{$d\gg p$}\\
   \cline{1-10}
   \multicolumn{2}{|c!{\vrule width 2pt}}{{Thermal conditions}}&$\lambda_T^{}\ll d$&$d\ll\lambda_T^{}\ll p$&$p\ll\lambda_T^{}$&$\lambda_T^{}\ll \{d,p\}$&$\{d,p\}\ll \lambda_T^{}$&$\lambda_T^{}\ll p$&$p\ll\lambda_T^{}\ll d$&$d\ll \lambda_T^{}$\\
   \hhline{|==========|}
   \multirow{1}*{Many-body}&intra-ensemble&no&no&yes&no&yes&no&yes&yes\\
   \cline{2-10}
   interaction (MBI)&inter-ensemble&no&no&yes&no&yes&no&no&yes\\
   \cline{1-10}
   \multirow{1}*{Near-field}&intra-ensemble&no&no&yes&no&yes&no&yes&yes\\
   \cline{2-10}
   effect (NFE) &inter-ensemble&no&yes&yes&no&yes&no&no&yes\\
   \cline{1-10}
   \multicolumn{2}{|c!{\vrule width 2pt}}{Regime; formulas for $G(p,d)$}& \multicolumn{3}{c|}{Rarefied regime; Eq.~(\ref{G_N})}&\multicolumn{1}{c|}{Eqs.~(\ref{G_0})}& \multicolumn{1}{c|}{Eq.~(\ref{Gt})}& \multicolumn{2}{c|}{dense regime; Eq.~(\ref{G_N_dense}) }&Eq.~(\ref{Gt})\\
   \cline{1-10}
  \end{tabular}
  \label{MBI_NFE_regime}
 \end{table*}

As can be seen from the Table~\ref{MBI_NFE_regime}, the conditions for the MBI effects are more strict than than for the near-field effects. Once the geometric lengths for the 2D nanoparticle ensembles have been determined, we can easily know from the Table~\ref{MBI_NFE_regime} if there are MBI effects and NFE or not. There are no MBI effects and NFE on RHT under the thermal conditions listed in the columns 1, 4 and 6. According to the thermal conditions listed in the rest columns 2, 3, 5, 7 and 8, we can determine that near-field effects exist. While MBI effects only exist under the thermal condictions listed in the columns 3, 5, 7 and 8. For the condition in the column 2, the inter-ensemble NFE exists and none of MBI effects exists. Physically, the MBI effect is one kind of near-field effects. MBI effects not always exist where the near-field effects exist. 

It's worthwhile to mention that Eq.~(\ref{Gt}) is the most general formula for the thermal conductance suitable for all conditions. Under certain conditions, we can efficiently simplify the calculation  of the thermal conductance between nanoparticle ensembles by using the following simplified equations (i.e., ~(\ref{G_N}),~(\ref{G_N_dense})) and Eqs.~(\ref{G_0}), which have been summarized in the Table~\ref{MBI_NFE_regime} for convenience.

\subsubsection{Rarefied regime}
\label{rarefied regime}

In this regime, each ensemble behaves like a gas body when considering heat exchange between the 2D nanoparticle ensembles. RHT between the two 2D nanoparticle ensembles in this regime is dominant by the nanoparticles in proximity. The general formula Eq.~(\ref{Gt}) for the thermal conductance $G(p,d)$ between the nanoparticle ensembles in this regime can be simplified as follows.
\begin{equation}
G(p,d)=N\times G_{2np}(d),
\label{G_N}
\end{equation}
where $G(p,d)$ is the radiative thermal conductance between the two nanoparticle ensembles with the separation $d$ and lattice spacing $p$, $N$ is the number of nanoparticles in each ensemble, $G_{2np}(d)$ is the thermal conductance between two isolated nanoparticles with the separation $d$, which can be calculated easily by Eq.~(\ref{Gij}). The thermal conductance between ensembles composed of a lot of nanoparticles can be easily obtained by the simplified Eq.~(\ref{G_N}) rather than by the general formula Eq.~(\ref{Gt}).

\subsubsection{Dense regime}
\label{dense regime}

In this regime, each ensemble behaves like a solid-dense body when considering heat exchange between two 2D nanoparticle ensembles. Nanoparticles in each ensemble is indistinguishable and have nearly the same contribution to thermal radiation with each other. In this regime, the general formula Eq.~(\ref{Gt}) for the thermal conductance $G(p,d)$ between two nanoparticle ensembles can be simplified as follows.
\begin{equation}
G(p,d)=N^2\times G_{2np}(d).
\label{G_N_dense}
\end{equation}
As compared to Eq.~(\ref{G_N}) used for the \emph{rarefied regime}, in the \emph{dense regime} thermal conductance between  two finite-size square-lattice nanoparticle ensembles in Eq.~(\ref{G_N_dense}) is $N^2 \times G_{2np}(d)$ rather than $N \times G_{2np}(d)$. The thermal conductance between ensembles composed of a lot of nanoparticles can be easily obtained by the simplified Eq.~(\ref{G_N_dense}) of the general formula Eq.~(\ref{Gt}).

\subsubsection{non-MBI regime}
\label{non-MBI regime}
In this regime, many-body interaction effects on RHT is negligible. Hence, the thermal conductance $G(p,d)$ between two ensembles is the simple the pair-wise sum of the thermal conductance for all possible $N^2$ pairs and, being purely additive, it neglects all possible MBI effects. Indeed $G(p,d)$ is defined as:
\begin{equation}
\begin{aligned}
G(p,d)=\sum_{pair-wise}^{N^2}G_{{\rm pair}},
\label{G_0}
 \end{aligned}
\end{equation}
where $G_{{\rm pair}}$ is the thermal conductance between the two particles of an isolated pair (one particle in L and the another in U) calculated using Eq.~(\ref{Gij}) and assuming the pair as completely isolated (i.e. all the other $2N-2$ particles are absent).

\subsubsection{MBI regime}
\label{MBI regime}
In this regime, due to the complex many-body interaction, the simplified Eq.~(\ref{G_0}) of the general formula Eq.~(\ref{Gt}) for the thermal conductance in the \emph{non-MBI regime} can't be used anymore. The thermal conductance must be calculated using Eq.~(\ref{Gt}) with the help of the exact transmission coefficient of Eq.~(\ref{transmission}), since no approximation is possible.

Till now, we have clearly identified four asymptotic regimes of RHT between 2D finite-size square-lattice nanoparticle ensembles in total, \emph{rarefied regime}, \emph{dense regime}, \emph{non-MBI regime} and \emph{MBI regime}, respectively. As can been seen in Fig.~\ref{regime_map_RHT} and Table~\ref{MBI_NFE_regime}, the \emph{MBI regime} covers a part of the \emph{dense regime} and the rest part of the \emph{dense regime} is covered by the \emph{non-MBI regime}. The Fig.~\ref{regime_map_RHT} combined with the Table~\ref{MBI_NFE_regime} can be easily applied to determine the regime of RHT between 2D finite-size square-lattice nanoparticle ensembles.

In columns 1, 2, 4 and 6, there is no MBI effect corresponding to the \emph{non-MBI regime} and the thermal conductance can be calculated directly by the simplified Eq.~(\ref{G_0}) rather than the general formula Eq.~(\ref{Gt}) treating all possible nanoparticle pairs as they were isolated in vacuum without any influence by other nanoparticles. The simple Eq.~(\ref{G_N}) can be easily applied to obtain the thermal conductance under such conditions listed in the columns 1, 2 and 3 corresponding to the \emph{rarefied regime}. Columns 1 and 2 belong to both \emph{rarefield regime} and \emph{non-MBI regime}, therefore, simplification of the calculation for the thermal conductance can go further by using Eq.~(\ref{G_N}) as compared to that by the Eq.~(\ref{G_0}) simplified from the general formula Eq.~(\ref{Gt}).  Condition in the column 6 belongs to both \emph{non-MBI regime} and \emph{dense regime}, where the efficient Eq.~(\ref{G_N_dense}) can be applied to simplify the calculation of the thermal conductance as compared to the simplified Eq.~(\ref{G_0}) for the \emph{non-MBI regime}.

It\textquotesingle s also worthwhile to mention that there are some other length scales that might influence the thermal behavior. (a) The size of the ensemble. When the size of the ensemble is large enough, the boundary effect on radiative heat transfer is negligible and finite ensemble of such large size starts mimic the infinite ensemble. The effect of the size of ensembles will be discussed at the end of the following Sec.~\ref{RHF_2D_ensemble}A in brief. (b) The resonance wavelength of the nanoparticles. (c) The length at which the contribution of the electric dipoles and magnetic dipoles to the radiative heat transfer is comparable. Since the inhibition and enhancement of the electric dipole contribution and magnetic contribution to NFRHT might occur at different length scales. The total thermal conductance is the result of the competition between the electric and magnetic dipoles.

\section{Radiative heat transfer between 2D finite-size square-lattice nanoparticle ensembles: \\ Numerical results}

\label{RHF_2D_ensemble}
In this section, we numerically investigate the RHT between 2D finite-size square-lattice nanoparticle ensembles. Several materials have been considered, e.g., metallic Ag, dielectric SiC and phase-change material VO$_2$. Regimes of RHT between 2D finite-size square-lattice nanoparticle ensembles under given conditions are identified numerically. Effect of the many-body interaction on the RHT between 2D finite-size square-lattice nanoparticle ensembles is analyzed as focus. In addition, effects of the phase change of material and lateral translation of the two parallel 2D ensembles on RHT are analyzed. Particle radius ($a$) is 20 nm. The separation distance between any two particles center to center in the particle ensemble is larger than 3$a$, which makes the dipole approximation valid \cite{Ben2011,Ben2019,DongPrb2017}.

\subsection{Regime of RHT between 2D finite-size square-lattice nanoparticle ensembles}
\label{regime_of_RHT}

\begin{figure*} [htbp]
\centerline {\includegraphics[scale=0.7]{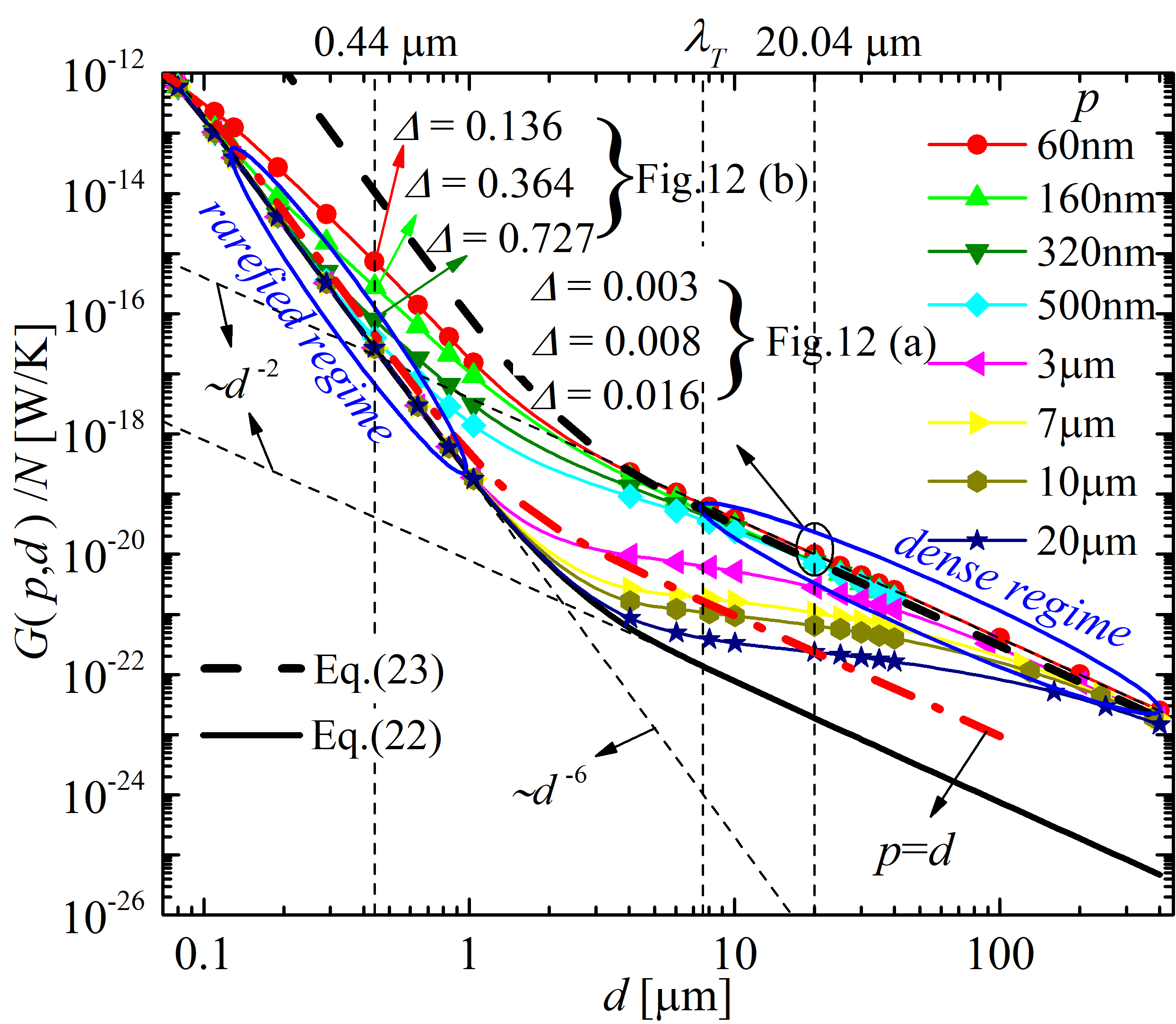}}
\caption{The scaled thermal conductance ($G(p,d)/N$) between 2D finite-size square-lattice nanoparticle ensembles as a function of separation distance $d$. We define $\varDelta =\frac{p}{d}$, and set $T=300$ K,  $a=20$ nm, $N=400$ and $p=60$ nm, 160 nm, 320 nm, 500 nm, 3 $\mu$m, 7 $\mu$m, 10 $\mu$m and 20 $\mu$m, respectively. Fitting lines of $d=0.44~\mu$m, $d=\lambda_T^{}$, $d=20.04~\mu$m and power law ($\sim d^{-2}$ and $\sim d^{-6}$) are added for reference. $G(p,d)/N$ calculated by the Eqs. (\ref{G_N}) and (\ref{G_N_dense}) as functions of $d$ are also added for reference. The thermal conductance $G(p,d)/N$ corresponding to $\varDelta=1~(d=p)$ is also added for reference. The data points used in the analysis on the thermal conductance spectrum in Fig.~\ref{periodicity} are marked for reference. Two regimes (\emph{rarefied regime} and \emph{dense regime}) are circled out in blue.}
\label{regime_RHT}
\end{figure*}

First of all, in order to investigate the regime of RHT between 2D finite-size square-lattice nanoparticle ensembles, we give a general description of the dependence of the scaled thermal conductance ($G(p,d)/N$) on the parameters $p$ and $d$, as shown in Fig.~\ref{regime_RHT}. Here we define the parameter $\varDelta =\frac{p}{d}$, and we set $T=300$ K, $N=400$, and $p=60$ nm, 160 nm, 320 nm, 500 nm, 3 $\mu$m, 7 $\mu$m, 10 $\mu$m and 20 $\mu$m, respectively. Lines corresponding to $d=0.44~\mu$m, $d=\lambda_T^{}$, $d=20.04~\mu$m and power law ($\sim d^{-2}$ and $\sim d^{-6}$) are added for reference. $G(p,d)/N \sim~d$ calculated by the Eqs. (\ref{G_N}) and (\ref{G_N_dense}) are also added for reference. The thermal conductance $G(p,d)/N$ corresponding to $\varDelta=1~(d=p)$ is also added for reference.

As shown in Fig.~\ref{regime_RHT}, when $d\ll \lambda_T^{}$ and $\varDelta \gg 1$ ($p \gg d$), the thermal conductance $G(p,d)/N$ follows the same power law $d^{-6}$ as that calculated by the Eqs. (\ref{G_N}). When $d \gg \lambda_T^{}$ and $\varDelta \ll 1$ ($p \ll d$), the thermal conductance $G(p,d)/N$ follows the same power law $d^{-2}$ as that calculated by the Eqs. (\ref{G_N_dense}). For a fixed separation $d$, the thermal conductance increases with the decreasing lattice spacing $p$. The nanoparticles inside each ensemble move from the far filed to the near field of its nearby nanoparticles when $p$ decreases from 20 $\mu$m to 60 nm. The near-field effect accounts for the increasing thermal conductance with decreasing $p$. As can be seen in Fig.~\ref{regime_RHT}, Eqs.~(\ref{G_N}) and~(\ref{G_N_dense}) give the lower and upper limits for the thermal conductance, respectively.

Then, based on the general description of the thermal conductance as show in Fig.~\ref{regime_RHT}, in order to numerically figure out the regime of RHT, a new parameter $\psi(p,d)$ is defined as follows.
\begin{equation}
\psi(p,d)=\frac{G(p,d)}{G_S(p,d)},
\label{IF}
\end{equation}
where $G(p,d)$ is the thermal conductance calculated by the general formula Eq.~(\ref{Gt}), $G_S(p,d)$ is the thermal conductance calculated by the simplified formulas of the general formula Eq.~(\ref{Gt}), i.e., Eqs.~(\ref{G_N}), ~(\ref{G_N_dense}) and ~(\ref{G_0}). The data points conrresponding to the \emph{rarefied regime} and \emph{dense regime} have been circled out in blue in Fig.~\ref{regime_RHT}.

\subsubsection{Rarefied regime}
\label{rarefied regime numeric}

\begin{figure} [htbp]
\centerline {\includegraphics[scale=0.45]{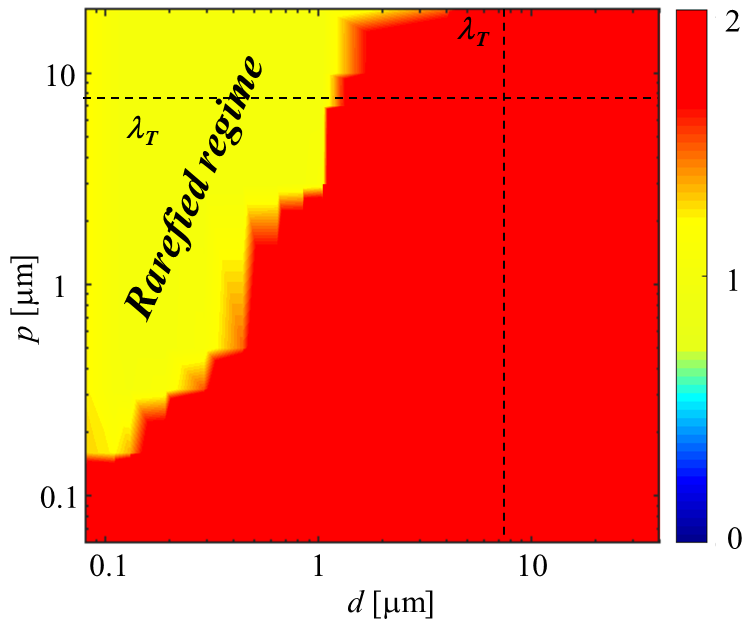}}
\caption{Contour of $\psi(p,d)$ calculated with the help of Eq.~(\ref{G_N}). The \emph{rarefied regime} is corresponding to the domain, where $\psi(p,d)\approx1$. $T=300$ K, $a=20$ nm. $N=400$.}
\label{rarefied_regime}
\end{figure}

In order to identify the \emph{rarefied regime}, the parameter $\psi(p,d)$ defined by Eq.~(\ref{IF}) is applied, where $G_S(p,d)$ is calculated by the simplified equation Eq.~(\ref{G_N}). We calculate the thermal conductance $G_S(p,d)$ in the considered domain ($p,d$) and give the contour of $\psi(p,d)$ in Fig.~\ref{rarefied_regime}. In the contour, the rarefied regime is corresponding to the region, where $\psi(p,d)\approx 1$. Under conditions that $T= 300$ K and $a=20$ nm, the region corresponding to the \emph{rarefied regime} is clearly identified in the Fig.~\ref{rarefied_regime}. In this region, the thermal conductance can be easily calculated by the simplified Eq.~(\ref{G_N}) of the general formula Eq.~(\ref{Gt}).

\subsubsection{Dense regime}
\label{dense regime numeric}
\begin{figure} [htbp]
\centerline {\includegraphics[scale=0.45]{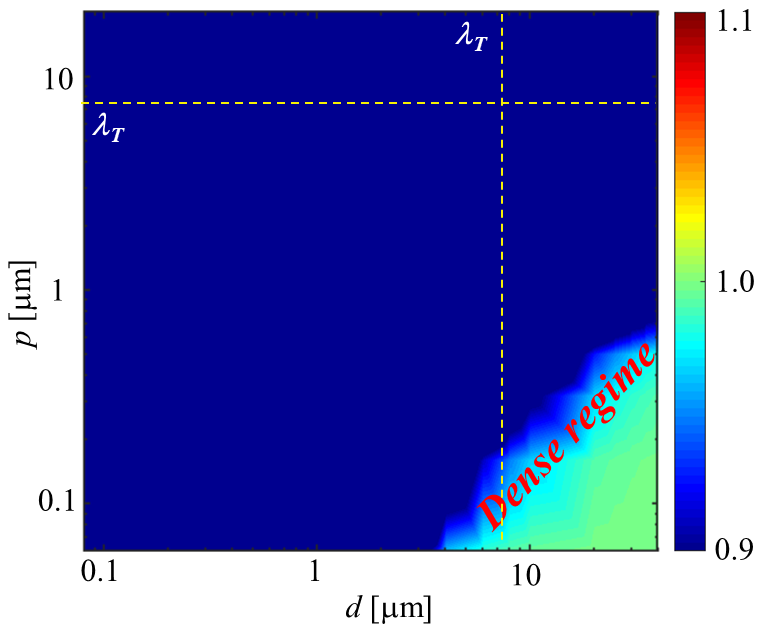}}
\caption{Contour of $\psi(p,d)$ calculated with the help of Eq.~(\ref{G_N_dense}). The \emph{dense regime} is corresponding to the domain, where $\psi(p,d)\approx1$. $T=300$ K, $a=20$ nm. $N=400$.}
\label{dense_regime}
\end{figure}

In order to identify the \emph{dense regime} clearly, the thermal conductance $G_S(p,d)$ calculated by Eq.~(\ref{G_N_dense}) is applied to calculate $\psi(p,d)$ in the considered domain $(p,d)$ and the contour of $\psi(p,d)$ is shown in Fig.~\ref{dense_regime}. In the contour, the \emph{dense regime} is corresponding to the region, where $\psi(p,d)\approx 1$. In this region, the thermal conductance between two emsembles composed of many nanoparticles can be easily calculated by the simplified Eq.~(\ref{G_N_dense}) of the general formula Eq.~(\ref{Gt}).

\begin{figure} [htbp]
     \centering
     \subfigure [Separation $d$ = 20 $\mu$m + 2$a$ for the far-field case] {\includegraphics[scale=0.4]{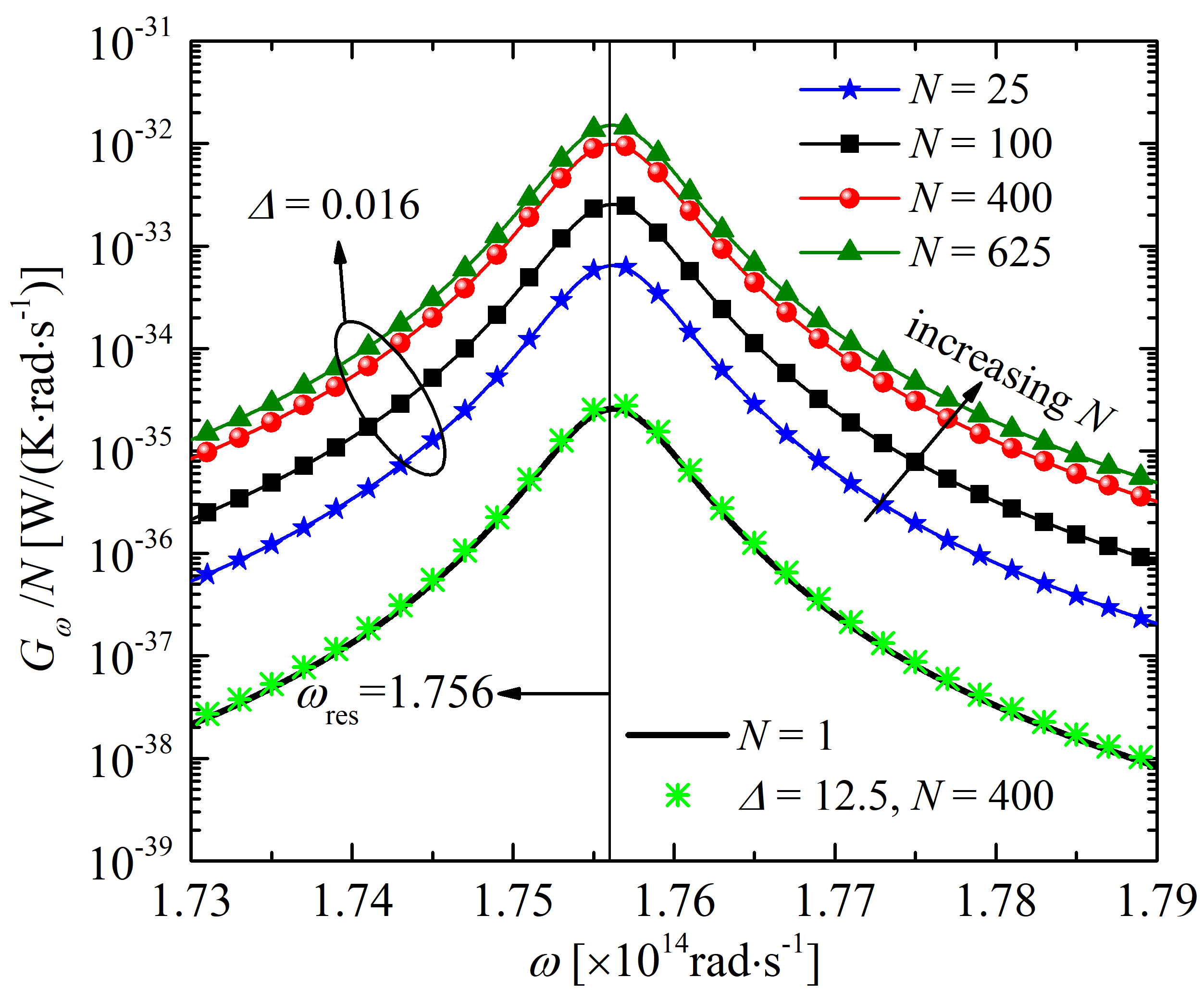}}\\
     \hspace{8pt}
     \subfigure [Separation $d$ = 40 nm + 2$a$ for the near-field case] {\includegraphics[scale=0.4]{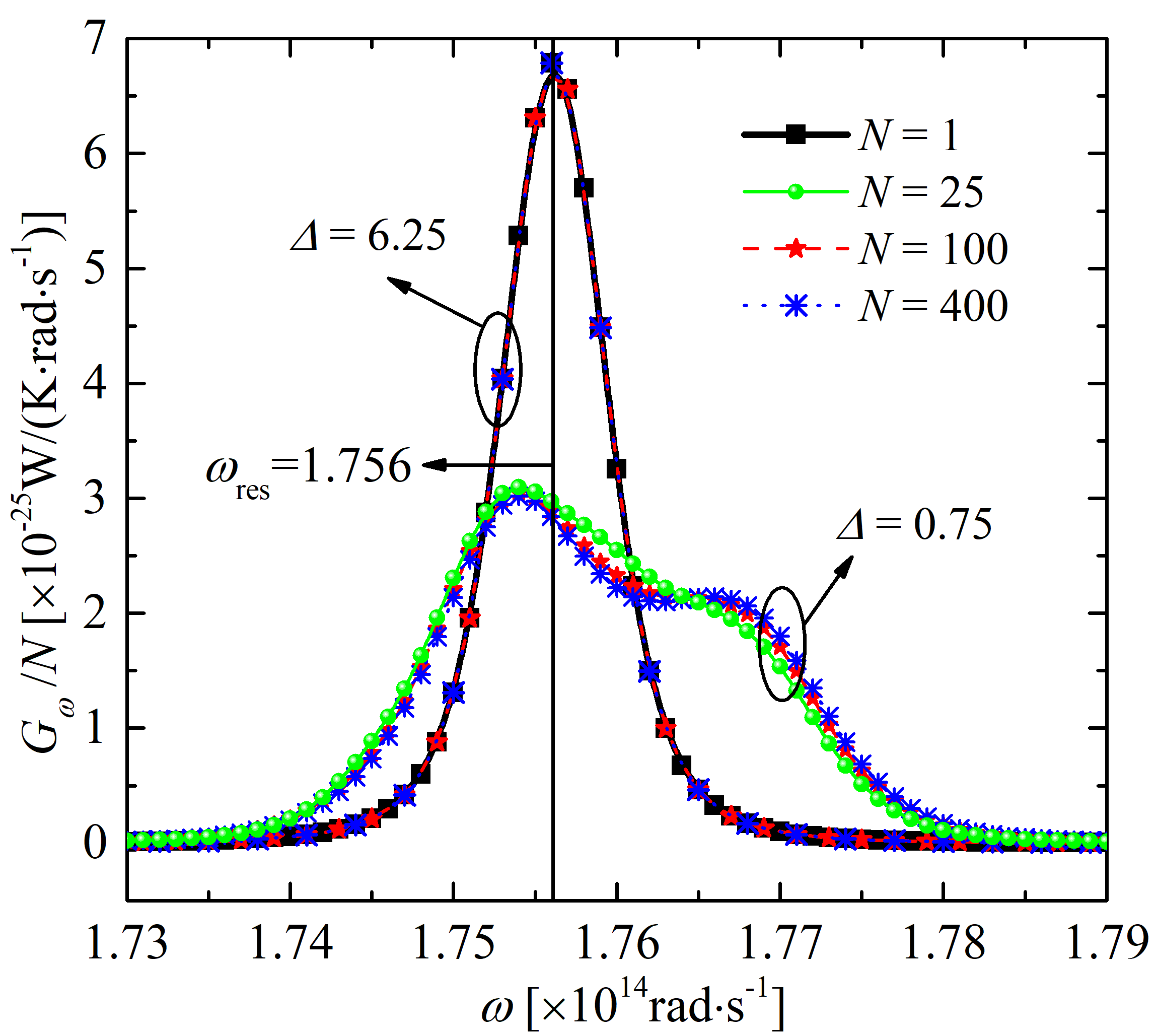}}
        \caption{The spectral thermal conductance $G_{\omega}^{}$ between SiC 2D finite-size square-lattice nanoparticle ensembles is scaled by the number of nanoparticles in each ensemble, $N$. Two cases are considered: (a) separation $d$ = 20 $\mu$m + 2$a$ between the two ensembles center to center for the far-field case, two cases have been considered, $\varDelta$ = 0.016 and 12.5, respectively. and (b) separation $d$ = 40 nm + 2$a$ for the near-field case, $\varDelta$ = 0.75 and 6.25. The thermal conductance spectrum between two nanoparticles separated by the same separation $d$ as that of the 2D ensemble is also added, which is the case $N = 1$ in the (a) and (b). The polarizability resonance frequency of the single SiC nanoparticle is added for reference, shown as $\omega_{\textrm{res}}^{}=1.756\times10^{14}$ rad$\cdot$s$^{-1}$.}
        \label{periodicity_limitation_case}
\end{figure}

In addition to the numerical identification for the \emph{rarefied regime} and \emph{dense regime} under the considered parameters for the ensembles, the effect of size of lattice on the RHT is also analyzed in both near field and far field. The spectral thermal conductance between two SiC 2D finite-size square-lattice nanoparticle ensembles separated by two different distances $d$ scaled by the amount of nanoparticles in each ensemble $N$ are shown in Fig.~\ref{periodicity_limitation_case}(a) and (b). The separation between the two 2D nanoparticle ensembles center to center are $d$ = 20 $\mu$m + 2$a$ (far-field case) and 40 nm + 2$a$ (near-field case), respectively. The thermal conductance spectrum between two isolated nanoparticles is also added for reference. The amount of nanoparticles in the 2D ensemble varies from case to case. The thermal conductance spectrum converges with increasing the number of nanoparticles in the 2D ensemble. Ensemble composed of 400 nanoparticles is sufficient to mimic the infinite 2D nanoparticle ensemble when considering heat exchange between the 2D ensembles.  As can be seen in Fig.~\ref{periodicity_limitation_case}(a) and (b), thermal conductance $G_{\omega}(p,d)/N$ reduces to that of two isolated nanoparticles when $\varDelta \gg 1~(p \gg d$) in both near field and far field, which is corresponding to the \emph{rarefied regime}.

\subsection{Effect of many-body interaction on RHT between 2D finite-size square-lattice nanoparticle ensembles}
\label{periodicity_effect_on_RHT}
Due to mutiple scattering in the ensemble composed of many particles, the many-body interaction has complex effect on RHT. Previous work focused on many-body interaction in particle system with different spatial arrangement (three-nanoparticle system \cite{Ben2011,Wang2016,Song2019}, 1D chain of nanoparticles \cite{Luo2019JQ}, 2D fractal nanoparticle ensemble \cite{Nikbakht2017}, clusters composed of hundreds of nanoparticles \cite{Dong2017JQ, Luo2019} and nanoparticles with a substrate \cite{DongPrb2018,Messina2018}), which can not only enhance  but also can inhibit NFRHT, and even have negligible effect on NFRHT. Here in this section, we focused on 2D square-lattice nanoparticle ensembles. Both quantitative and qualitative analyses on the many-body effect on NFRHT between 2D finite-size square-lattice nanoparticle ensemble are conducted.

\begin{figure} [htbp]
\centerline {\includegraphics[scale=0.45]{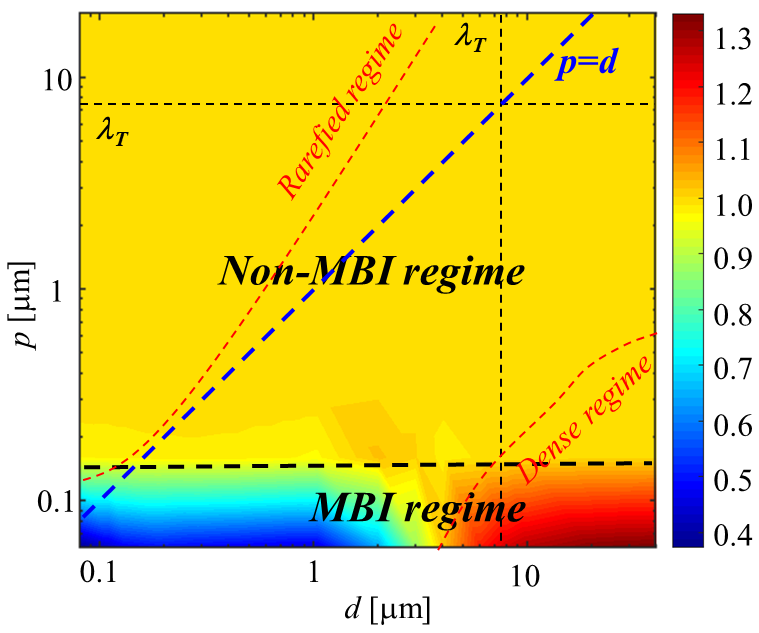}}
\caption{Contour of the ratio $\psi(p,d)$ for 2D SiC finite-size square-lattice nanoparticle ensembles as a function of $d$ and $p$. The black-bold-dash lines are the borderlines for the \emph{MBI regime}. The oblique blue-dash line corresponds to $p=d$. Lines for $p=\lambda_T^{}$ and $d=\lambda_T^{}$ are added. $a$ = 20 nm, $T$ = 300 K, $N$ = 400. The two red-dashed lines delimitate the \emph{rarefied regime} and \emph{dense regime} regions, already discussed in the Sec.~\ref{regime_of_RHT}.}
\label{regime_G_G0}
\end{figure}

The $\psi(p,d)$ defined by Eq.~(\ref{IF}) is applied to evaluate the many-body interaction quantitatively, where $G(p,d)$ is the thermal conductance between nanoparticle ensembles with the many-body interaction calculated by the general formula Eq.~(\ref{Gt}) and $G_S(p,d)$ is the thermal conductance between nanoparticle ensembles without the many-body interaction calculated by the simplified pair-wise summation formula Eq.~(\ref{G_0}). Here, the ratio $\psi(p,d)$ reflects the presence of MBI effects, and its numerical evaluation is shown in Fig.~\ref{regime_G_G0} (the parameters are listed in the caption of the figure). When $\psi(p,d)\approx 1$ (yellow region), the MBI is negligible. When $\psi(p,d)<1$ (blue region), the MBI inhibits the RHT. When $\psi(p,d)>1$ (red region), the MBI enhances the RHT. The oblique blue-dash line corresponds to $p=d$. Lines for $p=\lambda_T^{}$ and $d=\lambda_T^{}$ are added. The two red-dashed lines delimitate the \emph{rarefied regime} and \emph{dense regime} regions, already discussed in the Sec.~\ref{regime_of_RHT}.

By looking at the value of $\psi(p,d)$, two main regions can be identified, and separated with a black-bold-dash horizontal line $p\approx150~$nm in Fig.~\ref{regime_G_G0}: 1) the \emph{non-MBI regime}  (yellow region) corresponding to $\psi(p,d)\approx 1$  is essentially above that line, and 2) the  \emph{MBI regime} (blue/red regions) corresponding to $\psi(p,d)\neq 1$, is below that line. The \emph{MBI regime} only occupies a small part of the whole domain, where $p\ll \lambda_T^{}$. The rest large domain is corresponding to the \emph{non-MBI regime}, where MBI effects on RHT can be neglected safely and Eq.~(\ref{G_0}) is an excellent, time-saving, approximation of Eq.~(\ref{Gt}). If we focus on the MBI regime ($p\ll \lambda_T^{}$), we see that for $d<\lambda_T$ the MBI significantly inhibits the RHT, and  $\psi(p,d)$ reaches the minimum value $\psi(p,d)\approx 0.4$. For the region $d>\lambda_T$, on the contrary, MBI enhances the RHT, and $\psi(p,d)$ reaches its maximum value $\psi(p,d)\approx 1.4$. This enhancement effect is modest if compared to what happens between two nanoparticles due to the insertion of a third nanoparticle \cite{Ben2011}. In the MBI sector there is a transition region around $d=\lambda_T^{}$, where $\psi(p,d)\approx 1$.

\begin{figure} [htbp]
\centerline {\includegraphics[scale=0.4]{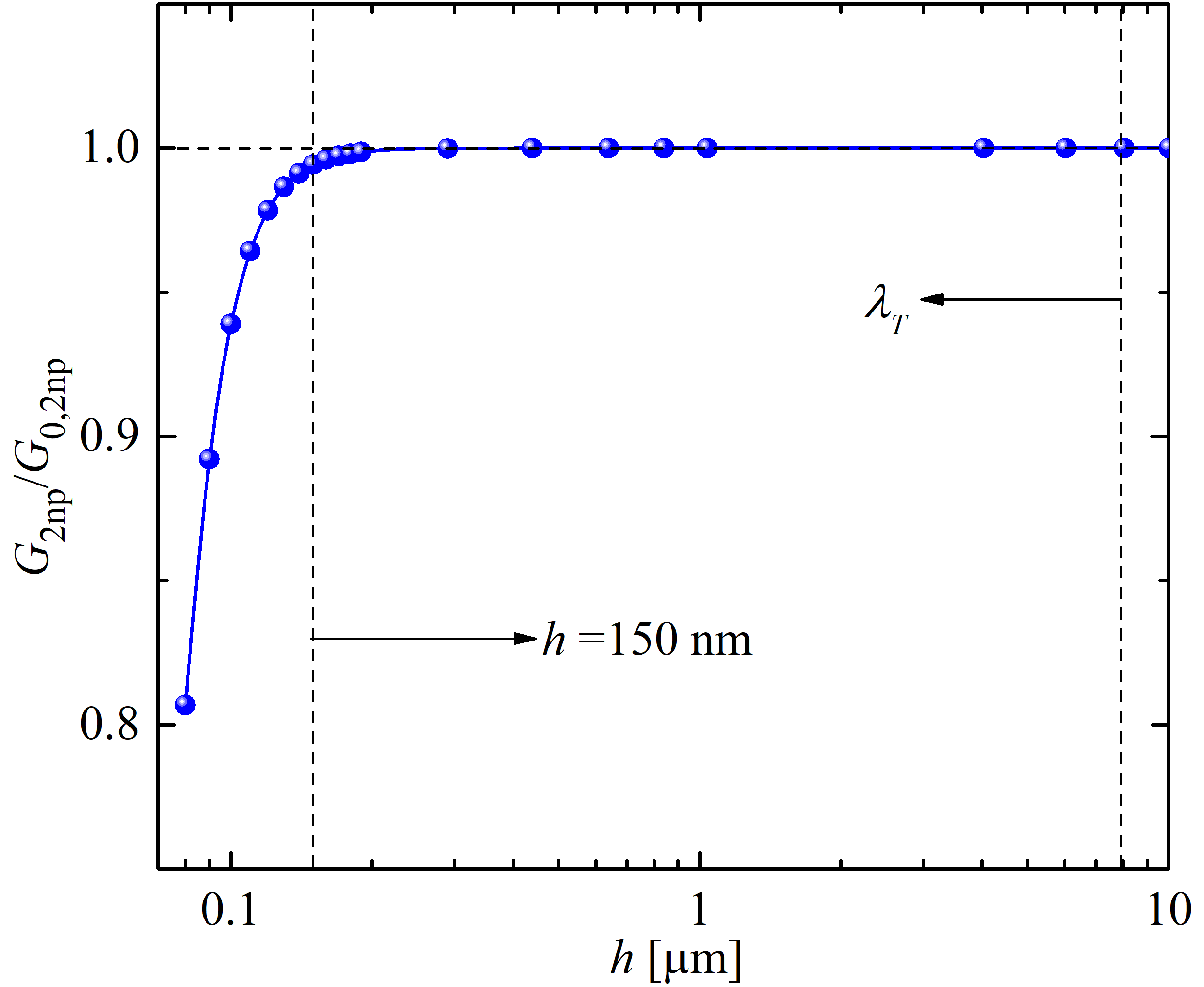}}
\caption{The ratio of the thermal conductance between two SiC nanoparticles with considering interaction between the two nanoparticles $G_{2np}$ by Eq.~(\ref{Gij}) to that without interaction between the two nanoparticles $G_{0,2np}$ as a function of separation $h$. $a=20$ nm, $T=300$ K. The lines corresponding to $h=150$ nm and $h=\lambda_T^{}$ are added for reference.}
\label{reflection_effect}
\end{figure}

In Fig.~\ref{regime_G_G0}, the horizontal line distinguishing the \emph{MBI regime} from the \emph{non-MBI regime} is at $p\approx150~$nm. We try now to understand the origin of this length scale. Let us start by investigating the multiple scattering occurring between the two nanoparticles of a single isolated pair. To this purpose, in Fig.~\ref{reflection_effect} we show the ratio of $G_{2np}$ to $G_{0,2np}$ as a function of the separation $h$, where  $G_{2np}$ is the complete thermal conductance between the two nanoparticles calculated using Eq.~(\ref{Gij}), while $G_{0,2np}$ is the thermal conductance neglecting the multiple scattering between the nanoparticles. The expression for $G_{0,2np}$ can be easily obtained from Eqs.~(\ref{ExchangedPower}) and (\ref{Gij}) with the help of the transmission coefficient in vacuum (see Eq. (30) of \cite{Luo2019}), which yields

\begin{equation}
\begin{aligned}
\mathcal{T}_{i,j}^{0}(\omega)=&\frac{4}{3}k^4\left[\textrm{Im}(\chi_E^i)\textrm{Im}(\chi_E^j)\textrm{Tr}(G_{0,ij}^{EE}G_{0,ij}^{EE\dagger})\right.\\&+\textrm{Im}(\chi_E^i)\textrm{Im}(\chi_H^j)\textrm{Tr}(G_{0,ij}^{EM}G_{0,ij}^{EM\dagger})\\&+\textrm{Im}(\chi_H^i)\textrm{Im}(\chi_E^j)\textrm{Tr}(G_{0,ij}^{ME}G_{0,ij}^{ME\dagger})\\&+\left.\textrm{Im}(\chi_H^i)\textrm{Im}(\chi_H^j)\textrm{Tr}(G_{0,ij}^{MM}G_{0,ij}^{MM\dagger})\right] ,
\label{transmission_0}
 \end{aligned}
\end{equation}
where $G_{0,ij}^{\nu\tau}$ ($\nu$,$\tau$=$E$~or~$M$) is the Green function in the free space, of which the explicit expression is given in Eqns.~(\ref{dielectric_tensor0})-(\ref{Compact_free_space_green_function}).
 In figure \ref{reflection_effect} the ratio $G_{2np}/G_{0,2np}$ calculated for SiC nanoparticles (radius $a=20$~nm, $T=300$~K) shows that scattering effects are relevant only for separations $h<L$, where $L\approx150$~nm is now a new length scale. When $L<h<\lambda_T\approx 7\mu$m, the system is in the near-field, while the multiple scattering is not relevant. We also note that the interaction between the two nanoparticles always inhibits the thermal conductance. We hence observe that the length scale $L$, setting the occurrence of multiple scattering inside an isolated pair, is compatibles with the value $p\approx$150nm setting the transition between non-MBI and MBI regions in Fig.~\ref{regime_G_G0}. From the same figure, se see that $L$ only affects the lattice spacing $p$, while has not signatures on the separation $d$ between the 2D planes. Hence, we deduce that the necessary condition to have MBI is the multiple scattering between the particles of the same plane. We also stress that the (less restrictive) near-field condition between in-plane particles ($p< \lambda_T$) is not sufficient to have MBI, and a multiple scattering (i.e.$p<L$) is needed.  Provided $p<L$, the MBI can also be present when the two 2D systems are in the far-field ($d>\lambda_T$).  It is also remarkable that  if $d<L$ but $\lambda_T> p>L$, we have multiple scattering between the particles of the opposite planes, but we do not have MBI.

\begin{figure} [htbp]
     \centering
     \subfigure [Separation $d$ = 20 $\mu$m + 2$a$ for the far-field case] {\includegraphics[scale=0.4]{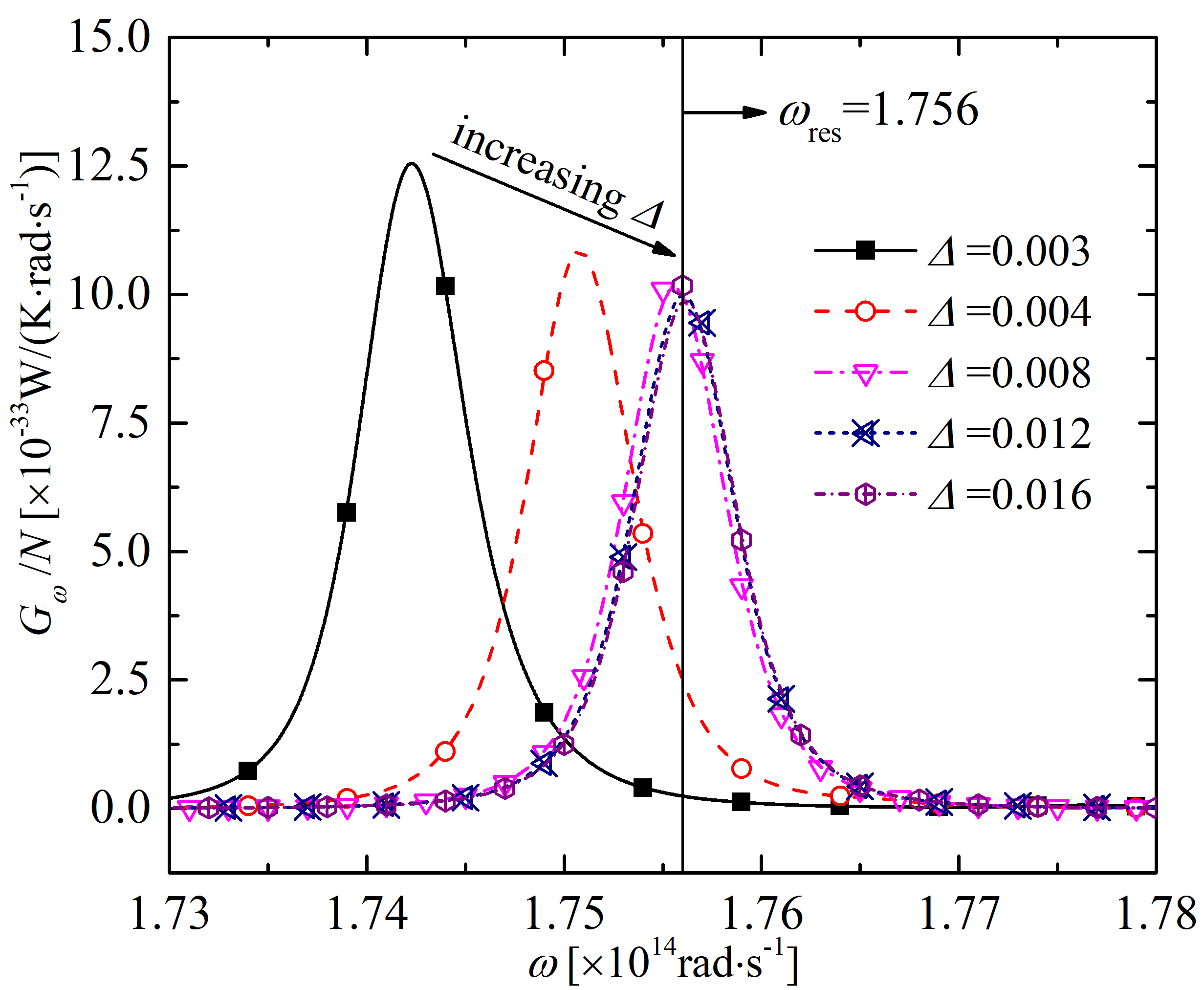}}\\
     \hspace{8pt}
     \subfigure [Separation $d$ = 400 nm + 2$a$ for the near-field case] {\includegraphics[scale=0.4]{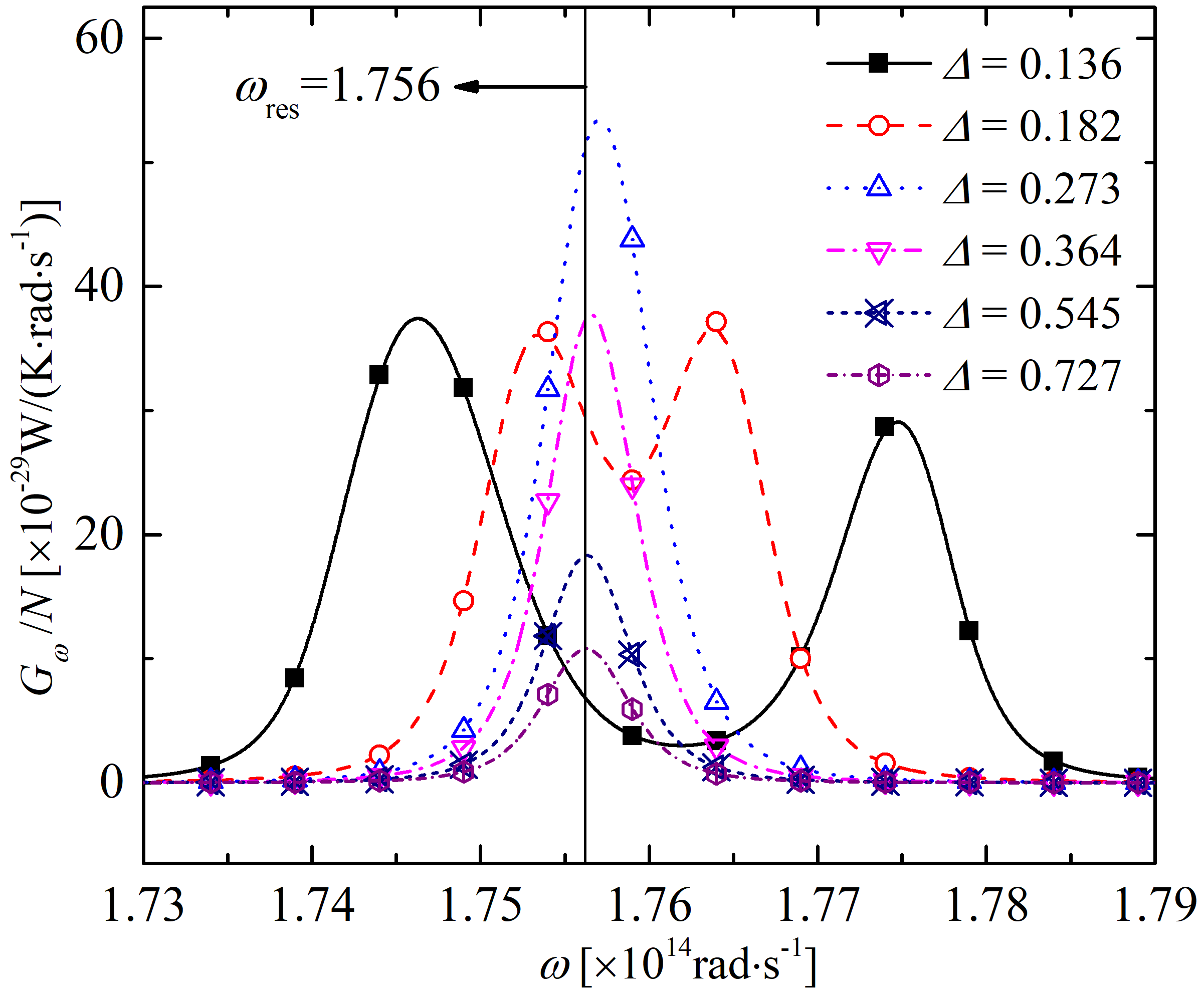}}
        \caption{The spectral thermal conductance $G_{\omega}^{}$ between two SiC particle ensembles with various $\varDelta=\frac{p}{d}$. Here $N=400$, $a=20$ nm, $T=300$ K. Two cases are considered: (a) separation $d$ = 20 $\mu$m + 2$a$ between the two particle ensembles center to center for the far-field case, $G_{\omega}^{}$ is scaled by the number of nanoparticles from one ensemble $N$ and (b) separation $d$ = 400 nm + 2$a$ for the near-field case, the $G_{\omega}^{}$ is scaled by the number $N$ of nanoparticles in each ensemble.}
        \label{periodicity}
\end{figure}

In order to better understand the different enhancing/inhibiting MBI effects observed in Fig.~\ref{regime_G_G0}, we consider below two cases for which we specifically analyse the spectral thermal conductance $G_{\omega}^{}(p,d)$ as shown in shown in Fig.~\ref{periodicity}:  (a) $d > \lambda_T~(d = 20~\mu$m + 2$a$, MBI effect enhances RHT) and (b) $d < \lambda_T~(d = 400$ nm + 2$a$, MBI effect inhibits RHT).

\subsubsection{\rm{Case 1: MBI enhances the RHT} $\{p\ll \lambda_T^{}\} \cap \{d>\lambda_T^{}\}$ }

In this case, corresponding to the red region of Fig.~\ref{regime_G_G0}, the MBI effect enhances the RHT. When two ensembles are separated by a large distance (far-field case, e.g., $d$ = 20 $\mu$m + 2$a$ $>\lambda_T$), the thermal conductance spectrum has only one peak as can be seen in Fig.~\ref{periodicity}(a) and the ratio $\psi(p,d)$ is near to but larger than 1. The maximum $\psi(p,d)$ in this case is less than 1.3. The value of the thermal conductance is less affected due to the weak intra-ensemble MBI effects inside each ensemble in this \emph{case 1}. The frequency corresponding with the peak of the thermal conductance spectrum shows a blue-shift behavior with the increasing $\varDelta$. In addition, the peak value of the spectral thermal conductance increases slightly with the decreasing $\varDelta$, which is corresponding to the slightly increasing thermal conductance with decreasing $p$ at a fixed $d$($= 20.04~\mu$m) observed in Fig.~\ref{regime_RHT}. In the \emph{Case 1}, the inter-ensemble MBI between the two ensembles is weak and can be neglected safely. However, the many-body interaction inside each of the ensemble (intra-ensemble MBI) is strong due to that nanoparticles from the same ensemble lie in the near field of each other. Therefore, the decreasing pure intra-ensemble MBI may account for the blue-shift of the peak frequency of the thermal conductance spectrum, when increasing periodicity $p$.

In addition, in the far-field case, the peak frequency of thermal conductance spectrum between 2D periodic ensembles approaches to the polarizability resonance frequency of the single particle (shown as $\omega_{\textrm{res}}^{}=1.756\times10^{14}$ rad$\cdot$s$^{-1}$ in Fig.~\ref{periodicity}), which satisfies $\Re(\epsilon(\omega))$ + 2 = 0. The spectral thermal conductance between two particle ensembles is the sum of the spectral thermal conductance of all nanoparticle couples, as can be seen from Eq.~(\ref{Gt}). The nanoparticle ensemble goes more dilute as $\varDelta$ increasing. Therefore, intra-ensemble MBI goes weaker with increasing $\varDelta$. Hence, the thermal conductance spectrum $G_{\omega}^{}(p,d)/N$ between 2D ensembles is similar to that between two isolated nanoparticles, where the intra-ensemble many-body interaction is negligible.

\subsubsection{\rm{Case 2: MBI inhibits the RHT} $\{p\ll \lambda_T^{}\} \cap \{d<\lambda_T^{}\}$ }

In this case, corresponding to the blue region of Fig.~\ref{regime_G_G0}, the MBI effect inhibits the RHT. When two ensembles are separated by a small distance (near-field case, e.g., $d$ = 400 nm + 2$a$ $<\lambda_T$), the thermal conductance spectrum is shown in Fig.~\ref{periodicity}~(b). The ratio $\psi(p,d) <1$ and MBI effect significantly inhibits the thermal conductance. The peak value of the thermal conductance spectrum increases heavily with decreasing $\varDelta$, which is also corresponding to that the scaled thermal conductance increases greatly with decreasing $\varDelta$ at a small $d$ ($0.44~\mu$m) observed in Fig.~\ref{regime_RHT}. In the \emph{Case 2}, only one peak of the thermal conductance spectrum can be observed for $\varDelta \geq 0.273$, where intra- and inter-ensemble MBI effects are weak. For $\varDelta = 0.136$ and 0.182, two spectral thermal conductance peaks can be observed, where the intra- and inter-ensemble MBI effects are strong. The co-existed strong intra- and inter-ensemble MBI effects may account for the two peaks of the thermal conductance spectrum between 2D finite-size square-lattice nanoparticle ensembles.

\begin{figure} [htbp]
\centerline {\includegraphics[scale=0.43]{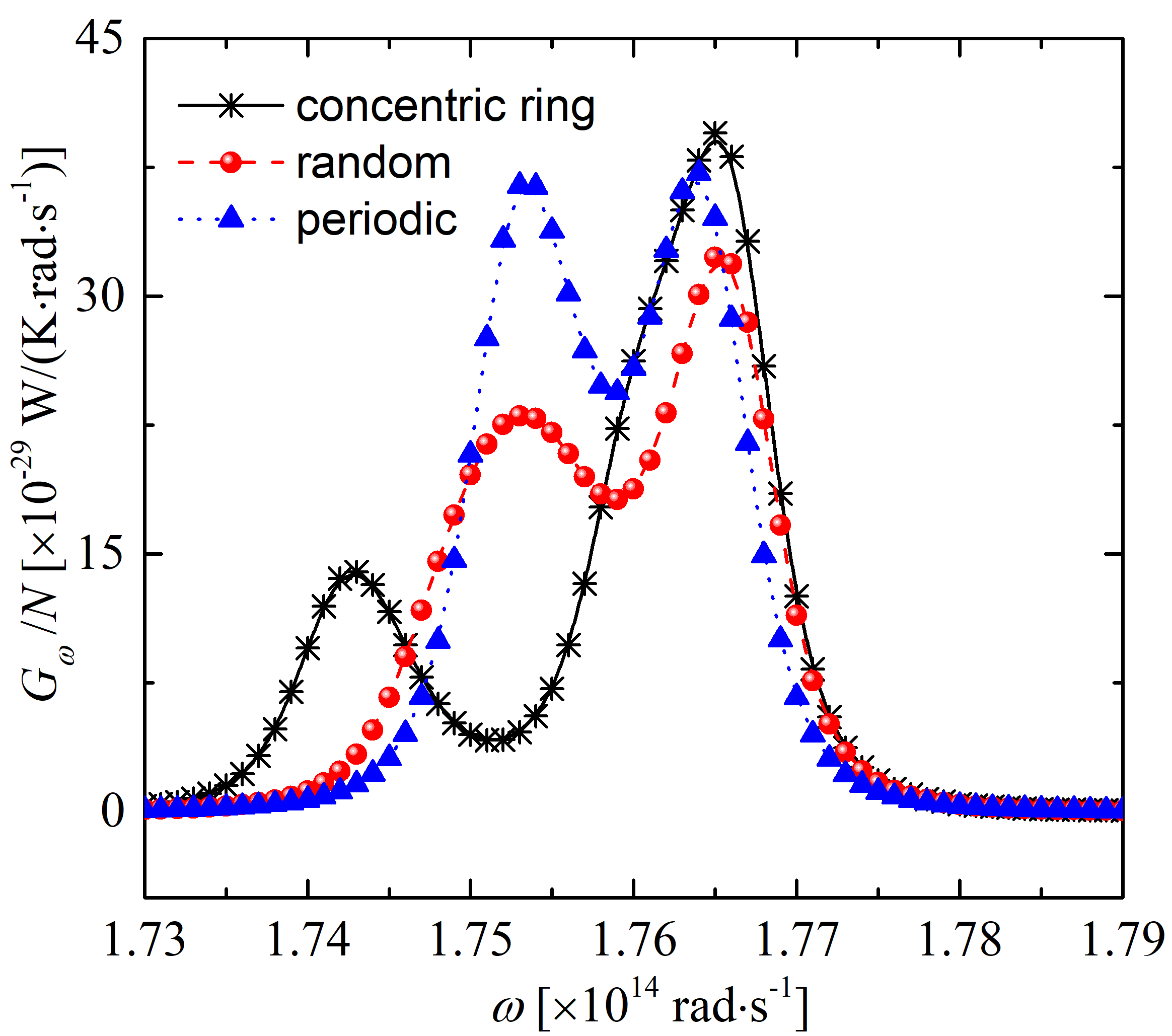}}
\caption{Thermal conductance spectrum between 2D ensembles with three kinds of particle distribution in dense regime: (a) periodic 2D ensemble ($\varDelta = 0.182$), (b) random 2D ensemble and (c) concentric ring configuration 2D ensemble. $a$ = 20 nm, $T$ = 300 K, $N$ =400 and separation $d$ = 400 nm + 2$a$ center to center.}
\label{peaks}
\end{figure}

An interesting question is if the phenomenon that thermal conductance spectrum between 2D nanoparticle ensembles in the \emph{Case 2} has two peaks is dependent on the particle distribution or not. Thermal conductance spectrum between 2D nanoparticle ensembles with three different kinds of the particle distribution is shown in Fig.~\ref{peaks}: (a) periodic 2D ensemble ($\varDelta = 0.182$), (b) random 2D ensemble and (c) concentric ring configuration 2D ensemble. Two peaks of the thermal conductance spectrum can be observed for all the three cases, which is independent on the particle distribution. That is to say the two peaks of thermal conductance spectrum between 2D nanoparticle ensembles with a small separation is due to the many-body interaction and independent on the particle distribution.

It\textquotesingle s also worthwhile to mention that the MBI is significantly dependent on the materials. We take the dielectric SiC as an example to analyze the NFRHT in the Secs.~\ref{RHF_2D_ensemble} A and B. In addition to the materials supporting resonance in the Planck\textquotesingle s window (e.g., SiC), we should also pay attention to the materials which do not support resonance in the Planck\textquotesingle s window (e.g., Ag). According to our previous works \cite{Luo2019,Luo2019JQ} for metallic Ag, due to the mismatch between the localized surface resonance wavelength and the thermal wavelength (Planck\textquotesingle s window), the MBI on RHT between metallic nanoparticle ensembles (e.g., clusters \cite{Luo2019} and nanoparticle chains \cite{Luo2019JQ}) can be safely neglected.

\subsection{Effect of the phase change on RHT between 2D finite-size square-lattice nanoparticle ensembles}
\label{Phase_change_effect}
The optical property of phase-change material below and above transition temperature is quite different from each other. The phase change of material has significant effect on RHT between planar surfaces in both near field and far field \cite{Taylor2017,Zheng2017,Ordonez-Miranda2018,Ordonez-Miranda2019,Szilard2019}. However, very few works on the effect of phase change on RHT between 2D nanoparticle ensembles, which is analyzed in this section.

\begin{figure} [htbp]
\centerline {\includegraphics[scale=0.4]{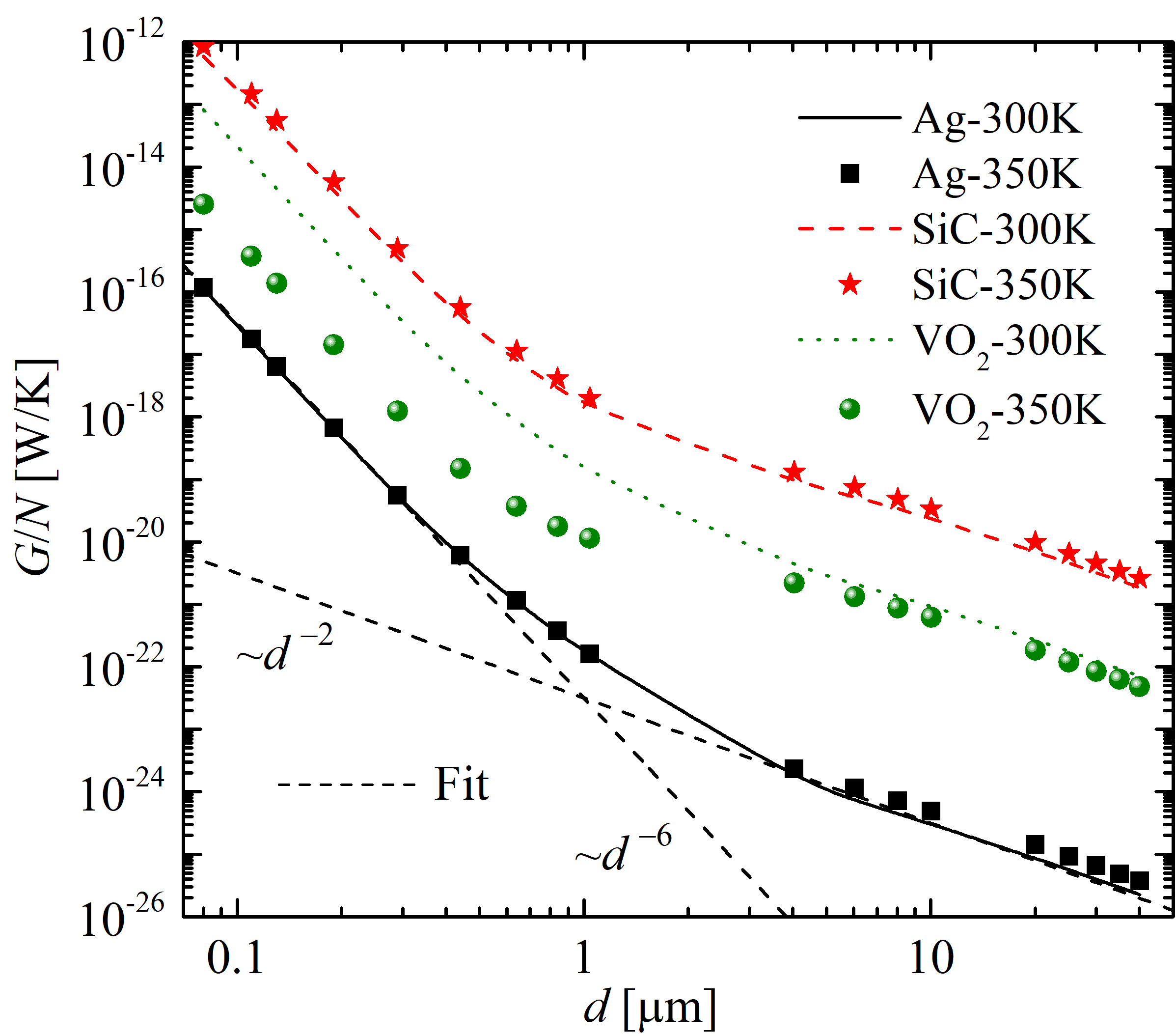}}
\caption{Thermal conductance between 2D periodic nanoparticle ensembles composed of metallic Ag, dielectric SiC and phase-change material VO$_2$. Two different temperatures below and above phase transition temperature of VO$_2$ have been considered, 300 K and 350 K, respectively. $a$ is 20 nm. $N$ is the number of nanoparticles in each ensembles L and U ($N$ = 400). $p$ = 500 nm. Two fitting lines of power law ($\sim d^{-2}$ and $\sim d^{-6}$) are added for reference.}
\label{material_effect}
\end{figure}

Thermal conductance between the 2D periodic nanoparticle ensembles as a function of the separation distance ($d$) is shown in Fig.~\ref{material_effect}. The symbols correspond to data obtained at $T=350$K and the lines correspond to data obtained at $T=300 $K. $N$ is the number of nanoparticles in each ensemble L and U ($N$ = 400). $p$ is the periodicity of the periodic ensemble ($p$ = 500 nm). Two fitting lines of power law ($\sim d^{-2}$ and $\sim d^{-6}$) are added for reference.

For the 2D periodic nanoparticle ensembles composed of metallic Ag or dielectric SiC, the thermal conductance at 350 K is similar to that at 300 K, which is also the case for 2D phase-change VO$_2$ ensembles with a large separation (the far-field case, large than 4 $\mu$m). However, for 2D periodic phase-change VO$_2$ nanoparticle ensembles with a small separation (the near-field case, smaller than 4 $\mu$m), thermal conductance at 300K is much higher than that at 350 K, which is an abnormal phenomenon different from the common sense that the thermal conductance often increases with temperature. Phase change of VO$_2$ significantly influences the RHT between 2D nanoparticle ensembles in the near field and has negligible effect on the RHT in the far field.

\begin{figure} [htbp]
\centerline {\includegraphics[scale=0.4]{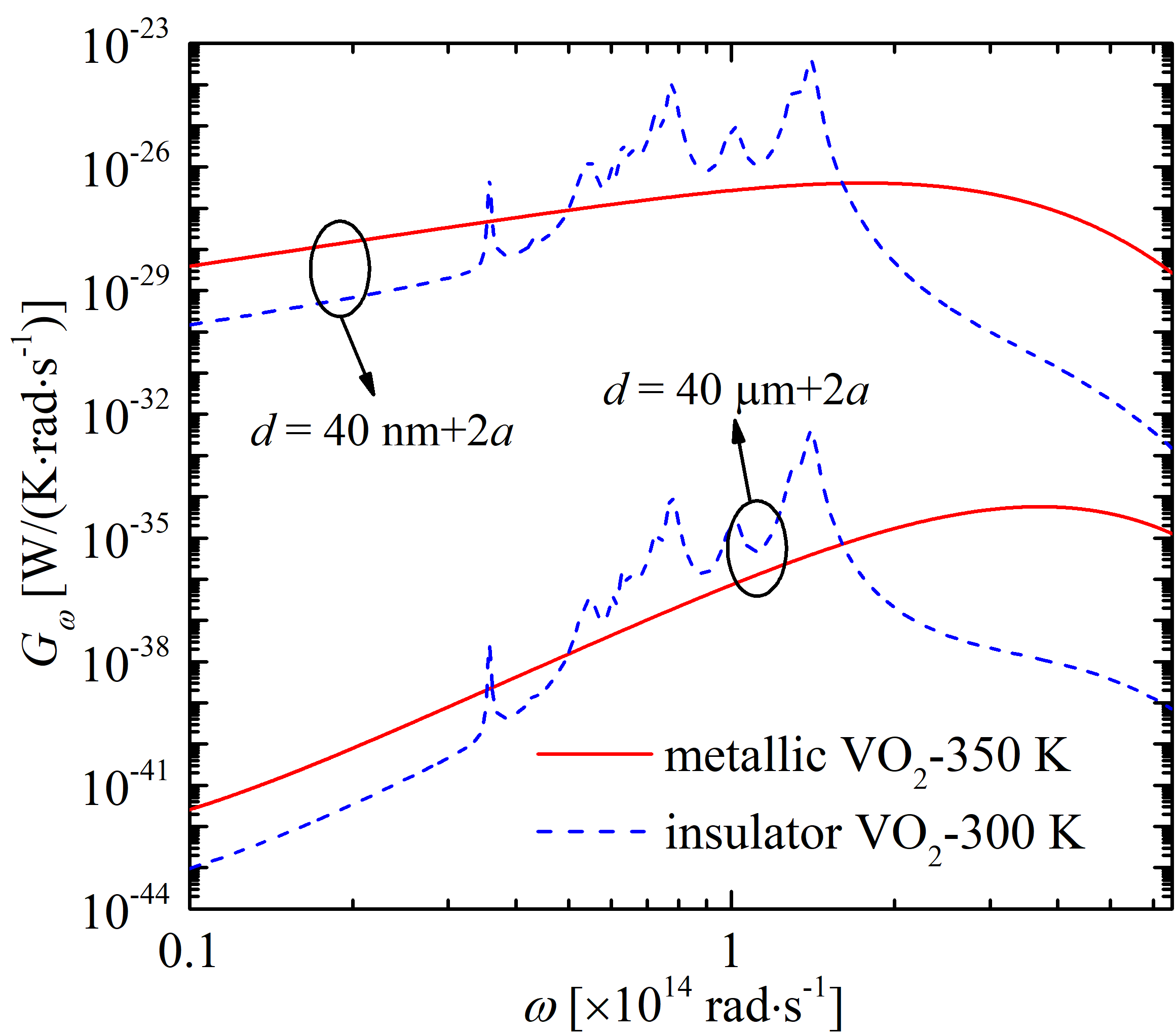}}
\caption{The spectral thermal conductance between two 2D VO$_2$ nanoparticle ensembles for $d~=~$40 nm + 2$a$ and 40 $\mu$m + 2$a$ at 300 K and 350 K. $a$ is 20 nm.}
\label{material_effect_spectrum}
\end{figure}

To understand insight of the abnormal phenomenon (mentioned in the 3$^{\textrm{rd}}$ paragraph) that the thermal conductance between two VO$_2$ nanoparticle ensembles decreases with increasing temperature, the spectral thermal conductances between VO$_2$ periodic particle ensembles at 300 K and 350 K are calculated at two different separation distances center to center ($d$ = 40 nm + 2$a$ and 40 $\mu$m + 2$a$ ), which are shown as Fig.~\ref{material_effect_spectrum}.  

As can be seen from the thermal conductance spectrum with $d$ = 40 nm at 300 K, there are two main and obvious peaks in the thermal conductance spectrum and the value of spectral thermal conductance around the peaks is much higher than that of the metallic-phase VO$_2$ nanoparticle ensembles at 350 K. The strong localized surface phonon resonance (LSPhR) may account for the high thermal conductance between the insulator-phase VO$_2$ particle ensembles at 300 K. As can be seen in Fig.~\ref{polarizability}(b), there is a mismatch between the characteristic thermal frequency and polarizability resonance frequency of metallic-phase VO$_2$ nanoparticle, which may account for the lower thermal conductance as compared to that of insulator-phase VO$_2$ nanoparticles. In the far field, e.g., $d$ = 40 $\mu$m, the weak near-field effect of both insulator-phase and metallic-phase VO$_2$ particle ensembles may account for the similar thermal conductance with each other. 

\subsection{Oscillatory-like features of the RHT with translation of the upper 2D finite-size square-lattice nanoparticle ensembles}
\label{oscillatory_feature}

\begin{figure} [htbp]
\centerline {\includegraphics[scale=0.35]{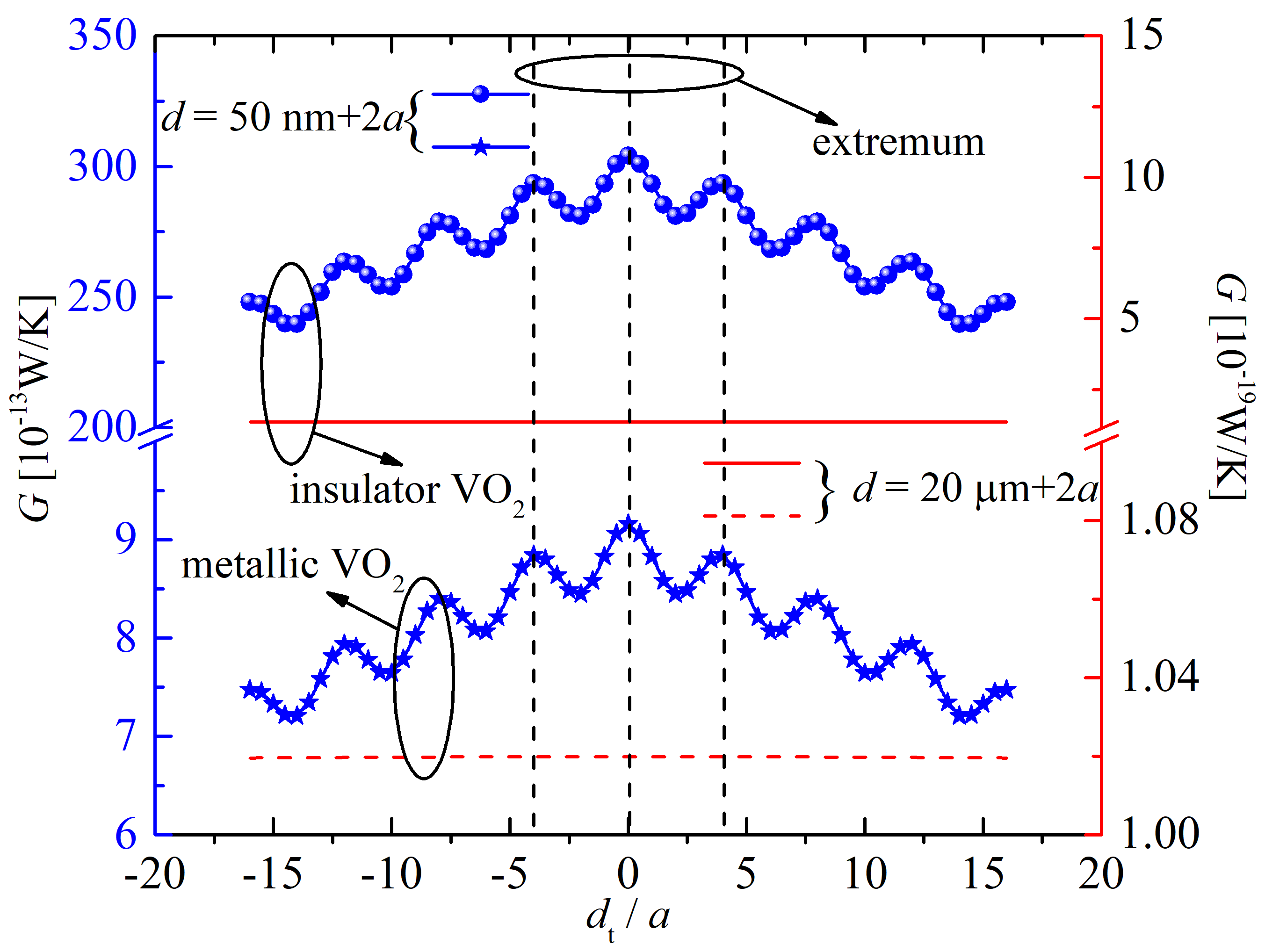}}
\caption{Thermal conductance between particle ensembles as function of the translation distance of one particle ensemble along its extending direction for several different cases. Red lines are for the data of metallic VO$_2$, while the blue ones are for the insulator VO$_2$. Two different separation distance between the two particle ensembles are considered, 50 nm + 2$a$ and 20 $\mu$m + 2$a$, respectively. $a$ is 20 nm. Periodicity of the 2D periodic nanoparticle ensemble $p$ is 80 nm. $N$ = 400.}
\label{thermal_coherent}
\end{figure}

A previous investigation on the NFRHT between two gold nanoparticle array layers with an extreme small separation of 1 nm layer edge to layer edge was reported \cite{Phan2013}, where the multipole contribution to NFRHT has be considered. An interesting oscillatory-like feature of the NFRHT with translation of one array along its extending direction was observed. However, the separation distance between 2D nanoparticle ensembles considered in this work is much larger than that considered in the reported work \cite{Phan2013}, where the dipole contribution dominates the NFRHT. Thermal conductance between nanoparticle ensembles as a function of the translation distance $d_\textrm{t}$ of one particle ensemble along the translation direction for two different separations $d$ are shown in Fig.~\ref{thermal_coherent}. Both insulator-phase and metallic-phase VO$_2$ nanoparticles are considered here. Two separation distances $d$ between the 2D periodic nanoparticle ensembles center to center are considered, 50 nm + 2$a$ and 20 $\mu$m + 2$a$, respectively. Nanoparticle radius $a$ is 20 nm. Periodicity of the 2D periodic nanoparticle ensemble $p$ is 80 nm.

\begin{figure} [htbp]
\centerline {\includegraphics[scale=0.35]{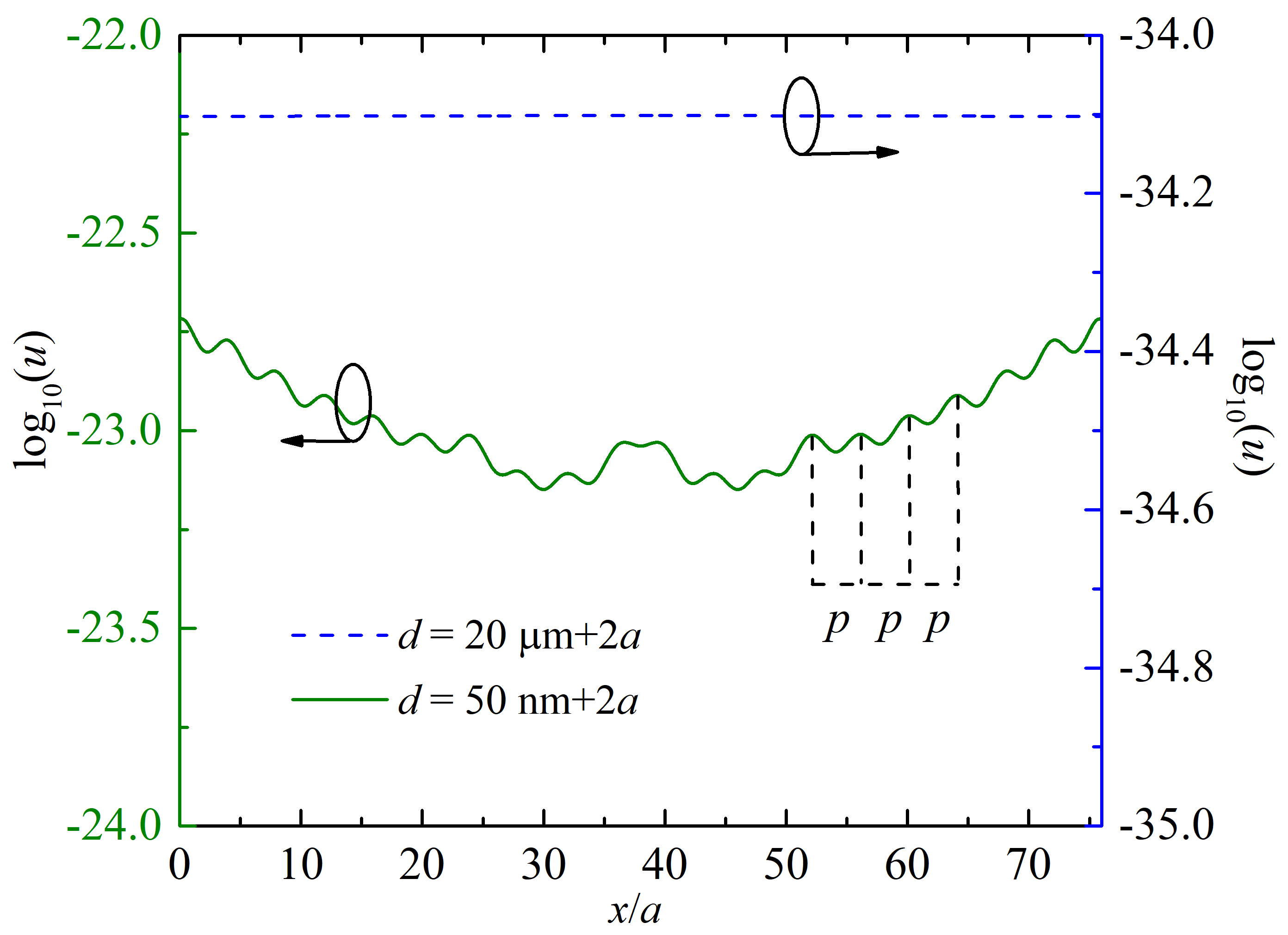}}
\caption{Energy density along the line with two different separation distances between the line and the lower insulator VO$_2$ nanoparticle ensemble at 300 K, 50 nm + 2$a$ and 20 $\mu$m + 2$a$, respectively. $p$ = 80 nm. $N$ = 400.}
\label{thermal_coherent_energy_density}
\end{figure}

When $d = 50~$nm + 2$a$, oscillatory-like features of RHT as a function of translation distance $d_\textrm{t}$ of the ensemble U relative to ensemble L is shown with star- and circle-symbols-lines in Fig.~\ref{thermal_coherent}. When $d = 20~\mu$m + 2$a$, no oscillatory-like phenomenon of RHT with translation distance can be observed, shown with solid lines in Fig.~\ref{thermal_coherent}. From Fig.~\ref{thermal_coherent}, the oscillatory periodicity is around 80 nm ($\sim p$), which is equal to the distance between the neighboring particles in the line parallel to the edge of the particle ensemble and is corresponding to the result observed in the reported work~\cite{Phan2013}. The oscillatory periodicity of thermal conductance with translation distance $d_\textrm{t}$ is similar to the spatial periodicity of the 2D periodic nanoparticle ensembles, which may not be an accident. It is noticed that the local energy density distribution has been demonstrated to be very useful to help understanding of physical mechanism of NFRHT, e.g., to analyze the emission of plate \cite{Joulain2005} and NFRHT among particles in the many-particle system \cite{DongPrb2017,Luo2019JQ}. To understand insight of the relation between the two kinds of periodicity, i.e., oscillatory periodicity of thermal conductance with translation distance and spatial periodicity $p$ of the 2D periodic nanoparticle ensemble, analysis of local energy density distribution is conducted. Energy density distribution above the insulator VO$_2$ nanoparticle ensemble is calculated along a line of interest (shown with the yellow line in Fig.~\ref{Structure_diagram}), which is shown in Fig.~\ref{thermal_coherent_energy_density}.

The energy density along the line of interest (shown as the yellow line in Fig.~\ref{Structure_diagram}) with two different separation distance between the line and the lower insulator VO$_2$ particle ensemble, $d = 50$ nm + 2$a$ and $d =$ 20 $\mu$m + 2$a$, respectively, are shown in Fig.~\ref{thermal_coherent_energy_density}. In the far field, i.e., $d = 20~\mu$m + 2$a$, the energy density stays constant along the line, as shown in Fig.~\ref{thermal_coherent_energy_density}. However, the energy density shows an oscillatory-like feature along the line in the near field, i.e., the separation distance $d = 50$ nm + 2$a$, of which the oscillatory periodicity is also around 80 nm ($\sim p$). The oscillatory periodicity of thermal conductance shown in Fig.~\ref{thermal_coherent} is corresponding to the oscillatory periodicity of the energy density as shown in Fig.~\ref{thermal_coherent_energy_density}. The extremum of the thermal conductance can be reached when the upper ensemble is translated with a distance, of which the value is an integer times of periodicity $p$, as shown in Fig.\ref{thermal_coherent} with the dash line. The strong near-field effect may account for the oscillatory-like feature of energy density and thermal conductance in the near field ($d$ = 50 nm + 2$a$). With the separation distance increasing, the near-field effect decreases gradually, which accounts for that thermal conductance and energy density have no oscillatory-like feature in the far field. 

\section{Radiative thermal energy emitted by periodic and non-periodic 2D nanoparticle ensemble}
\label{RTE_2D_ensemble}
\begin{figure} [htbp]
\centerline {\includegraphics[width=0.5\textwidth]{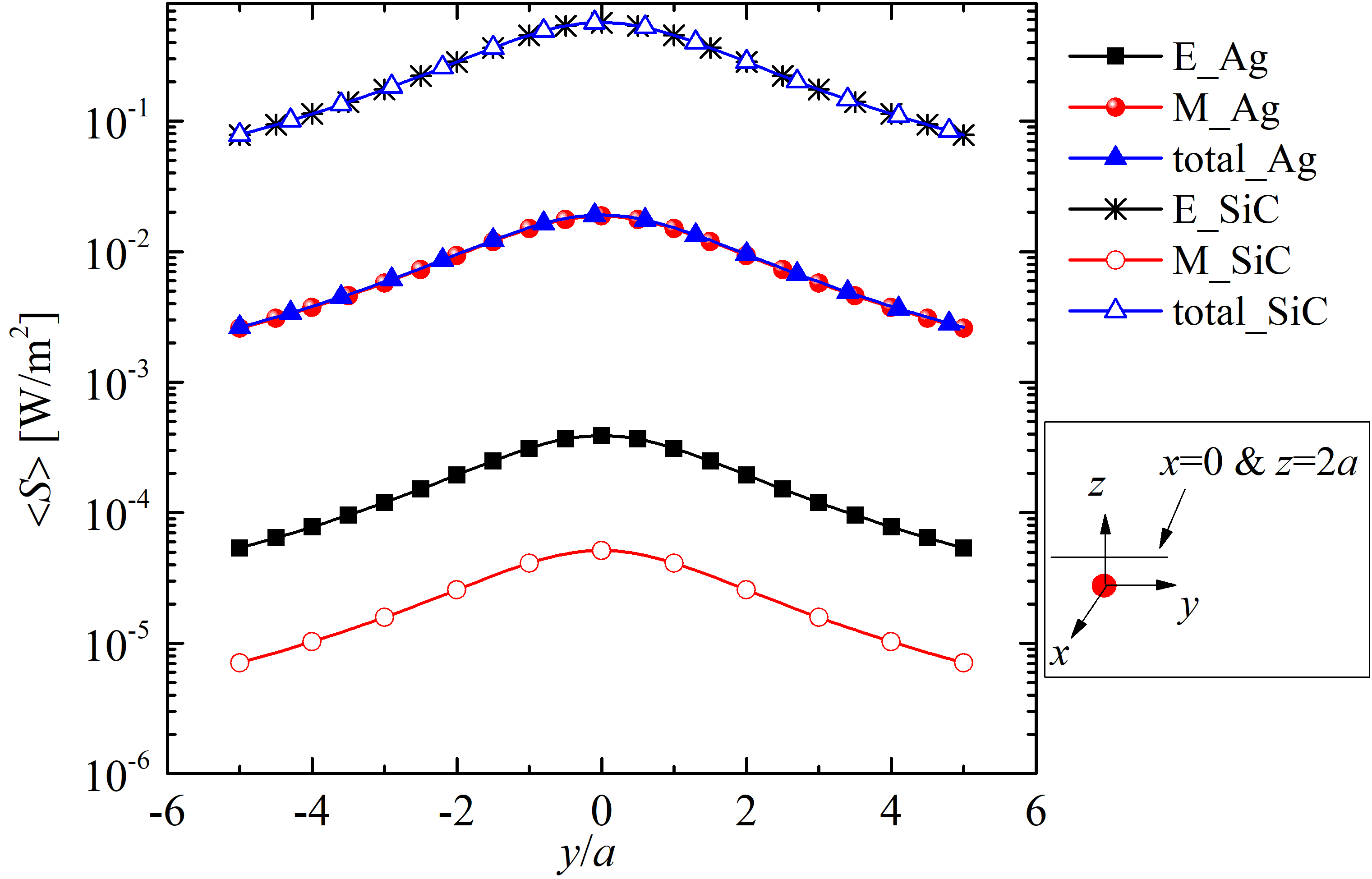}}
\caption{Poynting vector along the line of interest emitted by a nanoparticle due to electric (E) and magnetic (M) dipole contribution. The investigated physical model is shown in the inset. The nanoparticle is fixed at $(0,0,0)$. Both dielectric SiC and metallic Ag nanoparticles are considered. $T$ is 300 K. $a$ is 20 nm.}
\label{Poynting_EM}
\end{figure}

First of all, we compare the electric dipole and magnetic
dipole contribution to the radiative thermal energy emitted by a single nanoparticle. Total Poynting vector and Poynting vector contributed by the electric dipole (E) and magnetic dipole (M) emitted by a single nanoparticle is shown in Fig.~\ref{Poynting_EM}. Both metallic Ag and dielectric nanoparticles are considered. Particle radius $a$ is 20 nm. Temperature $T$ is 300 K. Poynting vector distribution along the line (defined by $x = 0 $~\&~$z = 2a$) shown with the inset of Fig.~\ref{Poynting_EM}. Both electric dipole and magnetic dipole contribute to the Poynting vector. For Poynting vector emitted by the dielectric SiC nanoparticle, the electric dipole contribution is four orders of magnitude larger than that of magnetic dipole contribution. While for Poynting vector emitted by the metallic Ag nanoparticle, the magnetic dipole contribution is two orders of magnitude larger than that of electric dipole contribution. It's worthy to mention that the magnetic dipole contribution dominates the radiative thermal energy emitted by metallic Ag nanoparticle. However, the electric dipole contribution dominates the radiative thermal energy emitted by the dielectric SiC nanoparticle.

\begin{figure} [htbp]
     \centering
     \subfigure [Poynting vector along the line with equation $y=0~\&~z=50$ nm] {\includegraphics[scale=0.42]{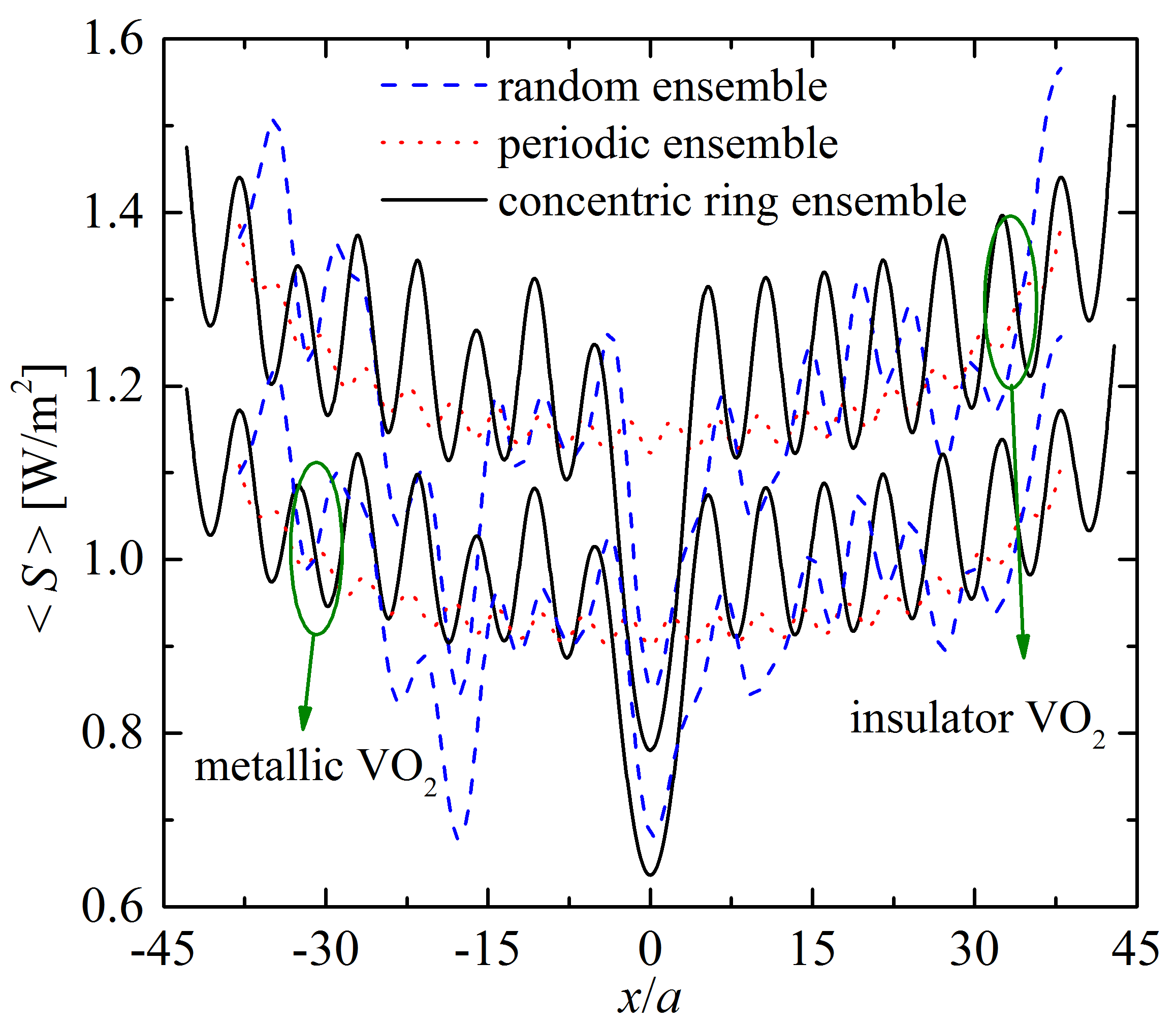}}\\
     \hspace{8pt}
     \subfigure [Poynting vector along the line with equation $y=0~\&~z=200$ nm] {\includegraphics[scale=0.4]{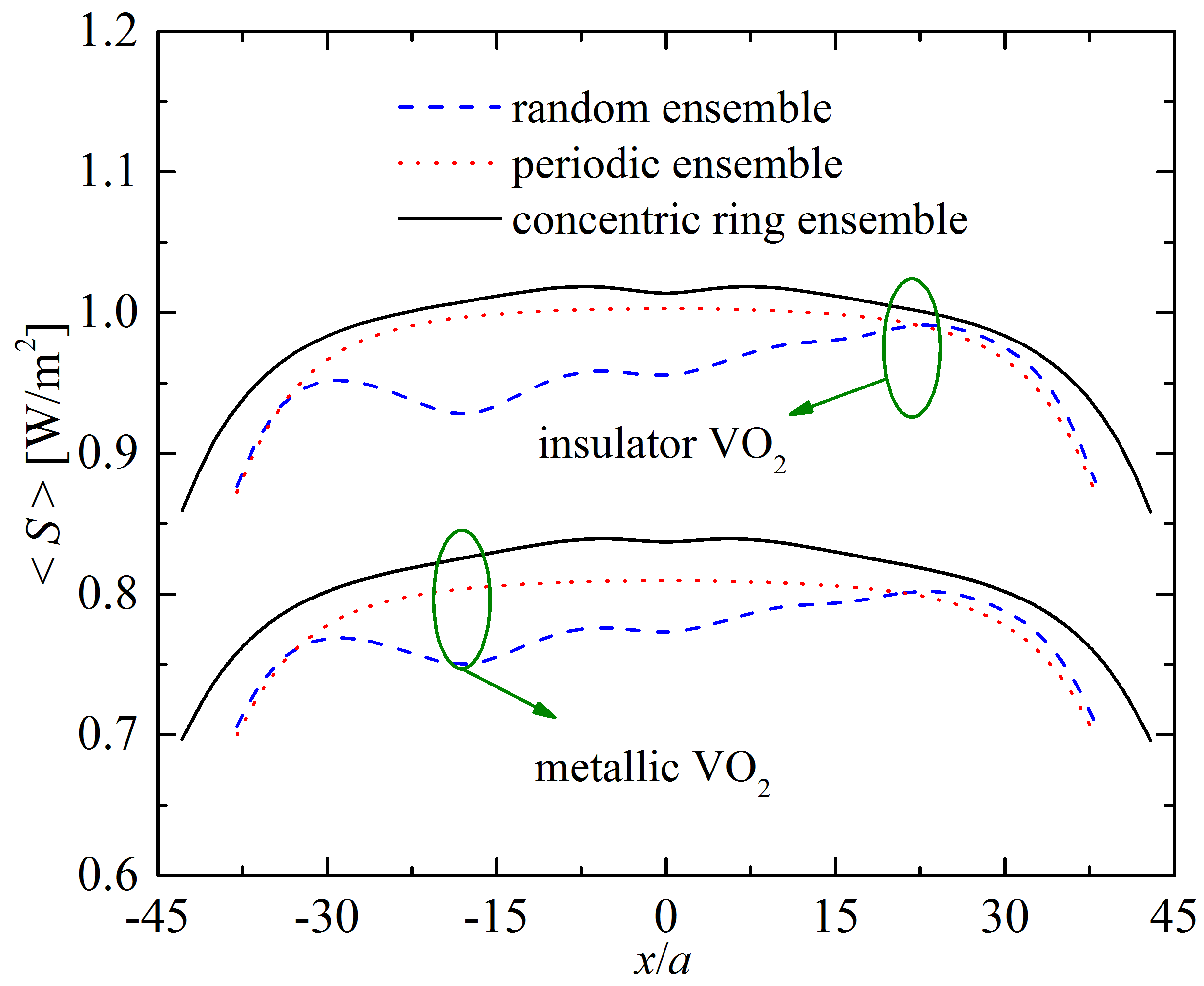}}
        \caption{The Poynting vector along the two lines (defined by $y=0~\&~z=50$ nm and $y=0~\&~z=200$ nm, respectively) for three different particle ensembles. Both insulator-phase VO$_2$ particle (300 K) and metallic-phase VO$_2$ particle (350 K) are considered. $a$ is 20 nm. $N$ = 400.}
        \label{Poynting}
\end{figure}

Then the effect of the nanoparticle distribution on the radiative thermal energy is analyzed. Three different kinds of nanoparticle distributions are considered here: (a) periodic 2D ensemble; (b) random 2D ensemble and (c) concentric ring configuration 2D ensemble, respectively, as shown in the Fig.~\ref{2dstructure}. Considering that the overlap influences the RHT significantly \cite{Phan2013}, all three different 2D particle ensembles are generated to have similar effective occupied area to each other. Nanoparticle radius $a$ is 20 nm. The Poynting vector along the two lines (defined by (1) $y=0~\&~z=50$ nm and (2) $y=0~\&~z=200$ nm, respectively) for all the three different particle ensembles are shown in Fig.~\ref{Poynting} (a) and (b). Both insulator-phase and metallic-phase VO$_2$ nanoparticles are considered. The number of nanoparticles $N$ in each ensemble is 400.

When the separation distance between the line and the ensemble is 50 nm, for periodic 2D ensemble, the Poynting vector shows an oscillatory feature, shown with the red dash line in Fig.~\ref{Poynting}(a). For the random 2D ensemble, no obvious regulation can be observed and fluctuation of the Poynting vector also can be observed. However, for concentric ring configuration 2D ensemble, near the center of the ensemble, the Poynting vector approaches to its minimum, which looks like a deep well. The geometrical configuration of concentric ring configuration particle ensemble may account for this focusing and inhibition of the radiative thermal energy. Meanwhile, the periodicity of the structure accounts for the oscillation of the Poynting vector near the periodic and concentric ring configuration 2D ensembles. The disordered nanoparticle distribution in the random 2D ensemble accounts for the irregular distribution of the Poynting vector. Wave effect of the thermally excited evanescent wave accounts for that the radiative thermal energy depends on the geometrical configuration.

When the separation between the line and 2D nanoparticle ensemble increases to 200 nm, as shown in Fig.~\ref{Poynting}(b), the Poynting vector along the line of interest is quite different from that observed in the case with 50 nm separation. No oscillatory feature of Poynting vector can be observed. The fluctuation of Poynting vector above the random 2D ensemble is much more dramatic than that of the other two 2D nanoparticle ensembles. The near-field effect decreases dramatically with the increasing separation from the source, which accounts for that Poynting vector in the near field is sensitive to the separation distance. Regularly 2D nanoparticle ensembles, e.g., periodic 2D ensemble and concentric ring configuration 2D ensembles, emit the radiative thermal energy in a similar way to each other. However, irregularity of the structure results in the irregular distribution and fluctuation of the radiative thermal energy.

From Fig.~\ref{Poynting}(a) and (b), it's noted that the phase change of $~$VO$_2$ only influences the value of the radiative thermal energy and has very weak effect on the distribution regulation of the radiative thermal energy. It\textquotesingle s worthwhile to mention that the sign of the curvature of diagrams has changed from Fig.~\ref{Poynting}(a) to Fig.~\ref{Poynting}(b). When the separation is small (Fig.~\ref{Poynting}(a)), the MBI is strong and inhibits the Poynting vector. Therefore, the Poynting near the center of the ensemble is inhibited heavily. However, when the separation is large (Fig.~\ref{Poynting}(b)), the MBI is weak and each nanoparticle emits energy separately and additionally. The Poynting vector near the center of ensemble is higher than that at other place.

\begin{figure} [htbp]
     \centering
     \subfigure [The averaged Poynting vector along the line with equation $y=0~\&~z=50$ nm] {\includegraphics[scale=0.4]{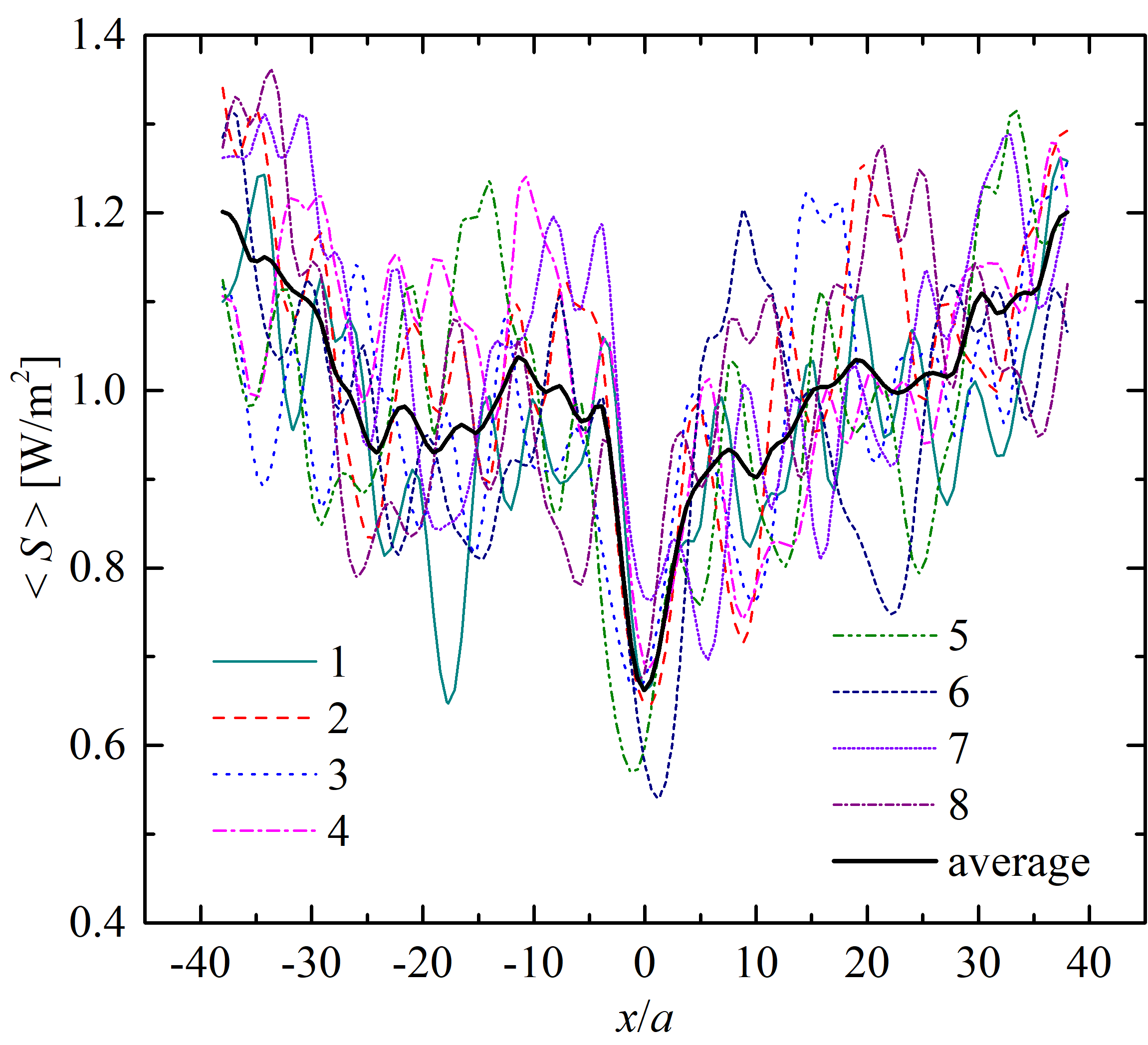}}\\
     \hspace{8pt}
     \subfigure [The averaged Poynting vector along the line with equation $y=0~\&~z=200$ nm] {\includegraphics[scale=0.4]{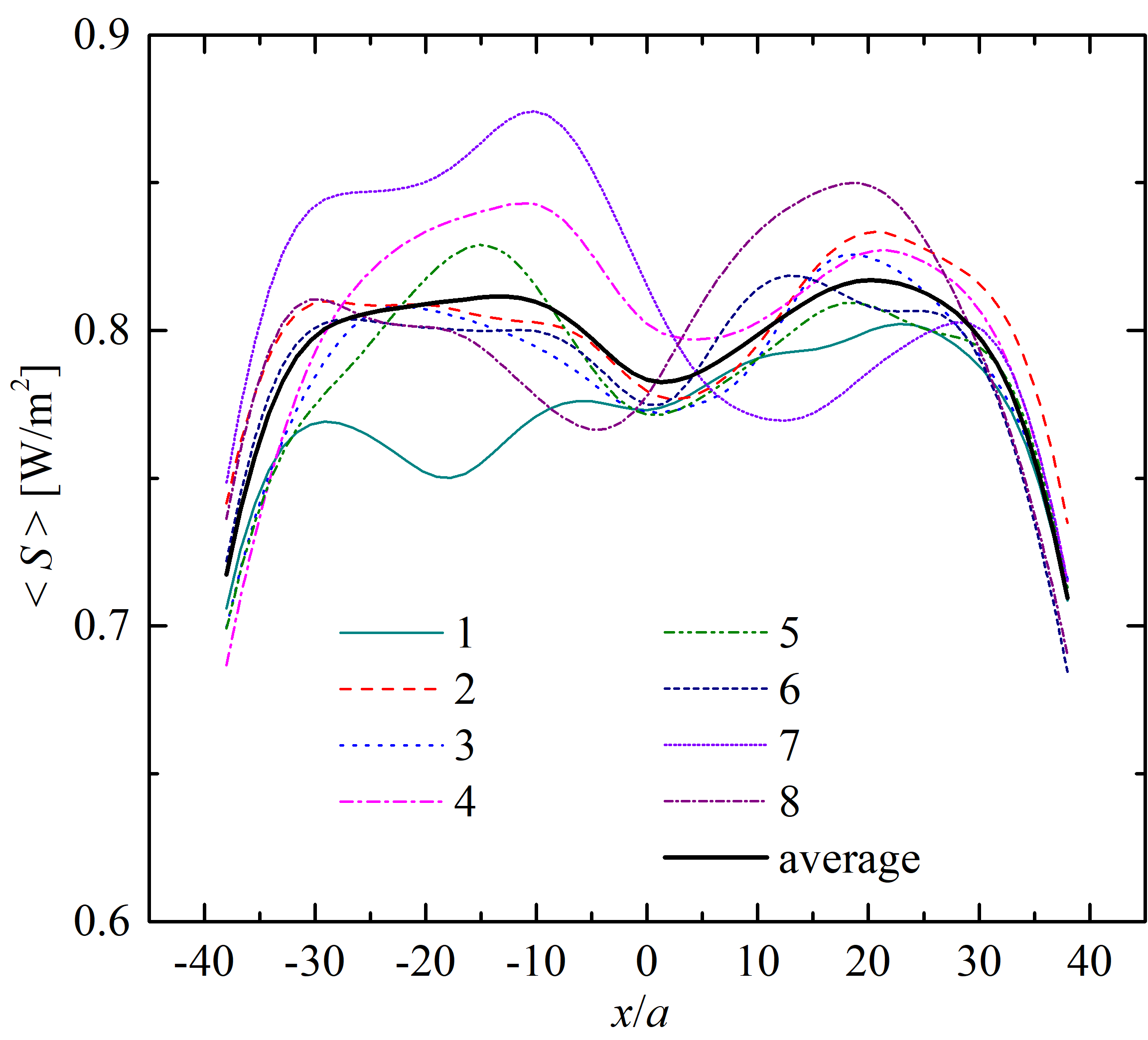}}
        \caption{The averaged Poynting vector over 8 random metallic-phase VO$_2$ nanoparticle ensembles along the two lines (defined by $y=0~\&~z=50$ nm and $y=0~\&~z=200$ nm, respectively). $T=$ 350 K. $a$ is 20 nm. $N$ = 400. The lines characterized by the number 1-8 are corresponding to the 8 different realizations of the random ensembles.}
        \label{S_average}
\end{figure}

Due to the statistical feature of the random ensemble, we average the Poynting vector over 8 realizations of random metallic-phase VO2 nanoparticle ensembles, as shown in Fig. ~\ref{S_average}. The Poynting vector along the two lines (defined by (1) $y=0~\&~z=50$ nm and (2) $y=0~\&~z=200$ nm, respectively) for the random ensembles of 8 realizations are also given in Fig.~\ref{S_average} (a) and (b). $T = 350$ K, $a=20$ nm. It\textquotesingle s shown that the average over the Poynting vector of 8 random ensembles reduces the asymmetry. Though each specific configuration of the random ensemble corresponding to a specific emitted Poynting vector distribution, average of the Poynting vector over a large number realization of random ensembles reduces the asymmetry of the data.

\begin{figure} [htbp]
\centerline {\includegraphics[scale=0.4]{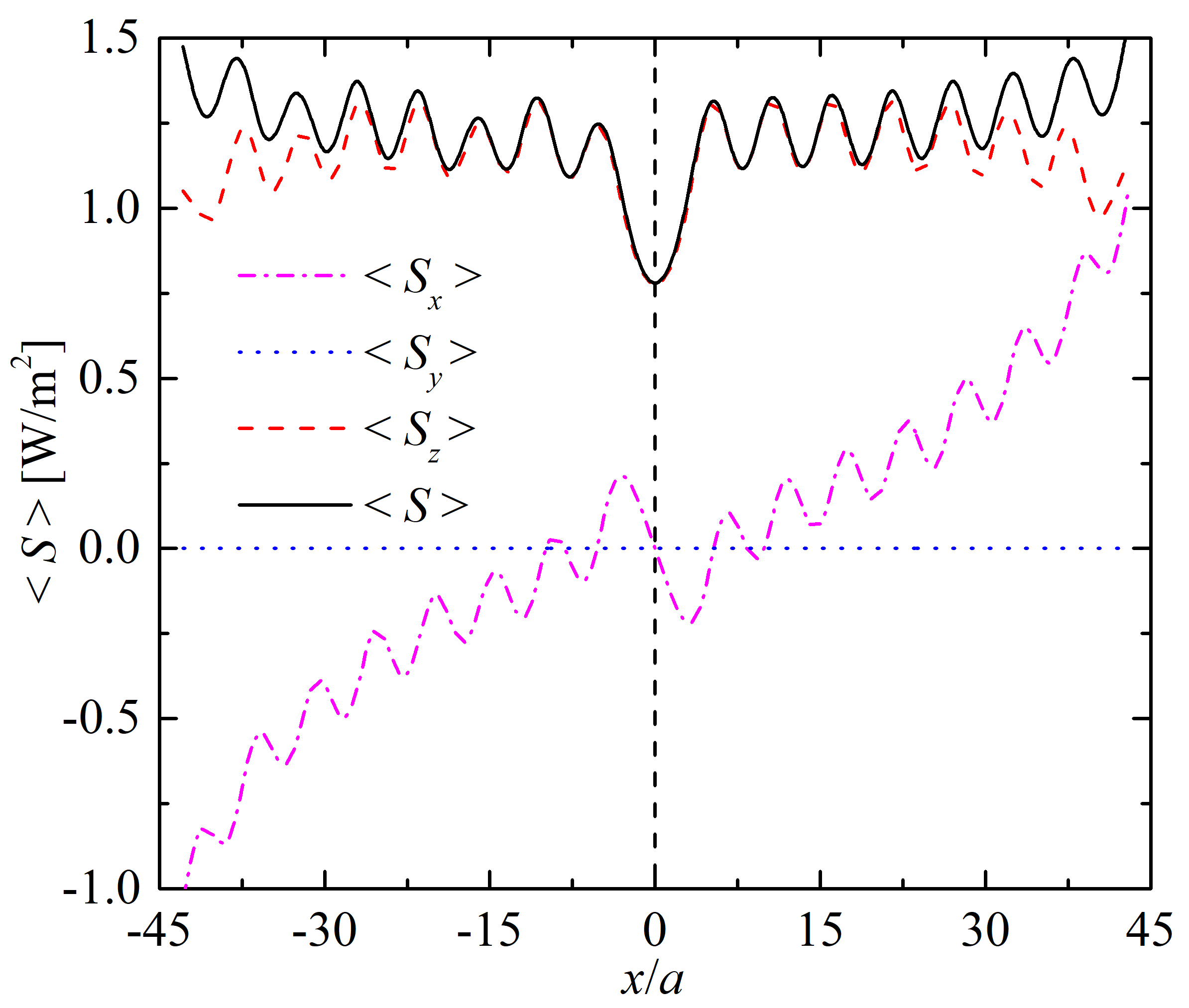}}
\caption{Components of the Poynting vector emitted by the concentric ring configuration 2D ensemble in $x$, $y$ and $z$ axes of the given Cartesian coordinate system along the line parallel to the 2D ensemble with 50 nm separation distance. $a$ = 20 nm. $N$ = 400.}
\label{Poynting_50nm_all_direction}
\end{figure}

Considering that the Poynting vector is a vector, it's worthy to point out its direction, which indicates the direction of the radiative thermal energy flow. All components of the Poynting vector emitted by the concentric ring configuration 2D ensemble along the line defined by $y=0~\&~z = 50$ nm are shown in Fig.~\ref{Poynting_50nm_all_direction}. The separation distance between the line of interest and the concentric ring configuration nanoparticle ensemble is 50 nm. Poynting vector component in the $y$ axis direction for the whole domain is zero. The radiative thermal energy flow in the plane perpendicular to the concentric ring configuration 2D ensemble. The symmetry of the concentric ring configuration nanoparticle ensemble geometrical configuration about the line of interest accounts for the radiative thermal energy flow in the plane perpendicular to the 2D ensemble. 

\begin{figure} [htbp]
     \centering{\includegraphics[scale=0.34]{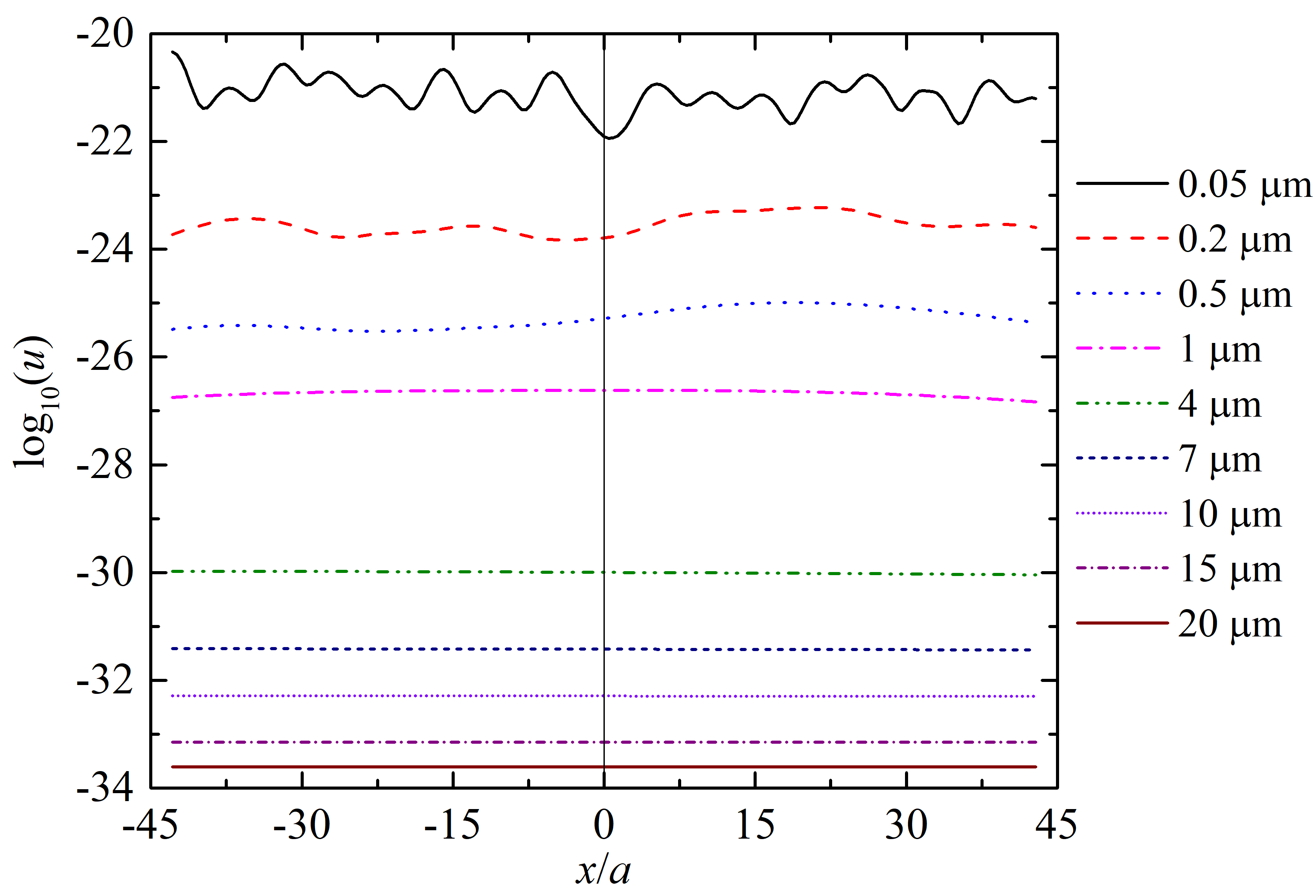}}
     \caption{Energy density along the line above the concentric ring configuration particle ensemble with 9 different separation distances, 0.05 $\mu$m, 0.2 $\mu$m, 0.5 $\mu$m, 1 $\mu$m, 4 $\mu$m, 7 $\mu$m, 10 $\mu$m, 15 $\mu$m and 20 $\mu$m, respectively. $N$ = 400.}
        \label{energy_density_Concentric ring-configuration particle ensemble}
\end{figure}

In addition, the radiative thermal energy emitted by the 2D nanoparticle ensemble is sensitive to the separation from the ensemble. To help for the understanding of this separation dependent radiative thermal energy, the energy density along the lines of interest (shown in Fig.~\ref{Structure_diagram}) parallel to the 2D ensemble with 9 different separations are shown in Fig.~\ref{energy_density_Concentric ring-configuration particle ensemble}. When the separation distance between the line and 2D nanoparticle ensemble is larger than 500 nm, the energy density does not vary from place to place any more. Howver, when the separation distance is less than 500 nm, the energy density oscillates along the line and is symmetrical in general about the center of the line, which can be attributed to the symmetry of the geometrical configuration of the 2D concentric ring configuration nanoparticle ensemble. Energy density is also dependent on the separation from the 2D ensemble, which is corresponding to that observed for the Poynting vector. The separation dependent near-field effect accounts for both of the separation dependent observation of the Poynting vector and energy density.

\section{Conclusion}
Radiative heat transfer (RHT) and radiative thermal energy concerning 2D nanoparticle ensembles is investigated in both the near field and far field by means of the CEMD approach and the proposed formulas of Poynting vector based on the framework of many-body radiative heat transfer theory. Asymptotic regimes of RHT between 2D finite-size square-lattice nanoparticle ensembles were summarized in the Table~\ref{MBI_NFE_regime} and the regime map Fig.~\ref{regime_map_RHT}. Four regimes and their corresponding thermal conductance formulas were given explicitly: (a) \emph{MBI regime} $\{p \ll \lambda_T^{}\}$, (b) \emph{non-MBI regime}, (c) \emph{rarefied regime}, and (d) \emph{dense regime}, respectively. In the \emph{rarefied regime}, \emph{dense regime} and \emph{non-MBI regime}, the thermal conductance formulas can be simplified as compared to the general formula.

According to the value of the parameter $\psi(p,d)$, we numerically identified the different asymptotic regimes in detail. In the \emph{MBI regime} $p \ll \lambda_T^{}$, MBI effects manifest themselves in different ways, depending on the separation $d$ significantly. In the \emph{MBI regime}, both value of the thermal conductance and shape of the thermal conductance spectrum $G_{\omega}$ can be influenced by the MBI effects. We considered two typical cases to analyze MBI effects on the spectrum $G_{\omega}$: \emph{Case 1} (MBI enhances the RHT, $\{p\ll \lambda_T^{}\} \cap \{d>\lambda_T^{}\}$) and \emph{Case 2} (MBI inhibits the RHT $\{p\ll \lambda_T^{}\} \cap \{d<\lambda_T^{}\}$). In the \emph{Case 1}, the decreasing pure intra-ensemble MBI accounts for the blue-shift of the peak of the spectrum $G_{\omega}$, and has a slight enhancement effect on the thermal conductance. In the \emph{Case 2}, the strong co-existed intra-ensemble and inter-ensemble MBI account for inhibition of RHT and the two peaks of the spectrum $G_{\omega}$ as compared to that of two isolated nanoparticles, which also has been demonstrated independent of the nanoparticle distribution.

The thermal conductance between 2D insulator-phase VO$_2$ nanoparticle ensembles with a small separation is much larger than that of the metallic-phase VO$_2$ nanoparticle ensembles due to the strong coupling in the insulator-phase VO$_2$ nanoparticle ensembles. However, this phase-change effect is negligible when the separation is large. An oscillatory-like feature of thermal conductance between 2D finite-size square-lattice nanoparticle ensembles with translation of the upper ensemble relative to the lower ensemble and thermally excited energy density distribution in the near field is observed, which is negligible in the far field. The strong near-field effect accounts for the oscillatory phenomenon of the thermal conductance and energy density. 

In addition, the formulas of the Poynting vector to evaluate the radiative thermal energy taking both electric and magnetic dipole contribution into consideration are given, which is an extension of the recent work ~\cite{Ben2019} and is applicable for not only dielectric but also metallic nanoparticle ensemble. It\textquotesingle s the magnetic dipole contribution that dominates the radiative thermal energy emitted by metallic nanoparticle. However, the electric dipole contribution dominates the radiative thermal energy emitted by the dielectric nanoparticle. The radiative thermal energy emitted by the 2D nanoparticle ensemble is sensitive to the particle distribution and the distance away from the ensemble. The RTE emitted by 2D concentric ring configuration nanoparticle ensemble has an inhibition feature near the ensemble center of the ensemble.

\begin{acknowledgments}
The support of this work by the National Natural Science Foundation of China (No. 51976045) is gratefully acknowledged. M.A. acknowledges support from the Institute Universitaire de France, Paris, France (UE). M.G.L. also thanks for the support from China Scholarship Council (No.201906120208).
\end{acknowledgments}

\appendix*
\section{Green\textquotesingle s function in many-particle system} 
\label{Dyadic_green_function}
In this Appendix, following the previous paper \cite{Luo2019}, we provide the details about how to calculate the many-body Green's function $G_{ij}^{\mu\nu}$ needed to evaluate the RHT of Eqs.~(\ref{power}) and (\ref{transmission}), and $G^{\nu\tau}(\mathbf{r},\mathbf{r}_i)$ needed to evaluate the RTE of Eqs.~(\ref{S_short}) and~(\ref{S_spectral_final}). Let us start by defining the Green\textquotesingle s function in free space:
\begin{equation}
G_{0}^{EE}(\textbf{r})=\frac{e^{ikr}}{4\pi r}\left[\left(1+\frac{ikr-1}{k^{2}r^{2}}\right)\mathbb{I}_{3}+\frac{3-3ikr-k^{2}r^{2}}{k^{2}r^{2}}\hat{\textbf{r}}\otimes\hat{\textbf{r}}\right],
\label{dielectric_tensor0}
\end{equation}
\begin{equation}
G_{0}^{ME}(\textbf{r})=\frac{e^{ikr}}{4\pi r}\left(1-\frac{1}{ikr}\right)\begin{pmatrix}
0 & -\hat{r}_{z} & \hat{r}_{y}\\
\hat{r}_{z} & 0 & -\hat{r}_{x}\\
-\hat{r}_{y} & \hat{r}_{x} & 0
\end{pmatrix},
\label{free_space_green_function}
\end{equation}
where $\mathbb{I}_3^{}$ is a $3\times3$ identity matrix, $r$ is the magnitude of the separation vector $\textbf{r}=\textbf{r}_f^{}-\textbf{r}_s^{}$ between the source point $\textbf{r}_s^{}$ and field point $\textbf{r}_f^{}$, $\hat{\textbf{r}}$ is the unit vector $\textbf{r}/r$ and $\hat{r}_{\nu=x,y,z}$ denotes its three Cartesian components, $\otimes$ denotes outer product of vectors. $G_{0}^{EM}(\textbf{r})=-G_{0}^{EM}(\textbf{r})$ and $G_{0}^{MM}(\textbf{r})=G_{0}^{EE}(\textbf{r})$. The Green\textquotesingle s function in free space can be written in a compact form as:
\begin{equation}
\mathbb{G}_{0,ij}^{}=
\begin{pmatrix}
\mu_{0}^{}\omega^{2}G_{0,ij}^{EE} & \mu_{0}^{}\omega G_{0,ij}^{EM}\\
k\omega G_{0,ij}^{ME} & k^{2}G_{0,ij}^{MM}
\end{pmatrix},
\label{Compact_free_space_green_function}
\end{equation}
where $G_{0,ij}^{\mu\nu}\equiv G_{0}^{\mu\nu}(\textbf{r}_i-\textbf{r}_j)$.
Also the many-body Green's function $G_{ij}^{\mu\nu}$ we are looking for can be written by exploiting the compact form 
\begin{equation}
\mathbb{G}_{ij}^{}=\begin{pmatrix}
\mu_{0}^{}\omega^{2}G_{ij}^{EE} & \mu_{0}^{}\omega G_{ij}^{EM}\\
k\omega G_{ij}^{ME} & k^{2}G_{ij}^{MM}
\end{pmatrix}.
\label{Compact_system_green_function}
\end{equation}
The function $\mathbb{G}_{ij}^{}$ is the solution of the equation
\begin{equation}
\begin{pmatrix}
0 & \mathbb{G}_{12}^{} & \cdots & \mathbb{G}_{1N}^{}\\
\mathbb{G}_{21}^{} & 0 & \ddots & \vdots\\
\vdots & \vdots & \ddots & \mathbb{G}_{(N-1)N}^{}\\
\mathbb{G}_{N1}^{} & \mathbb{G}_{N2}^{} & \cdots & 0
\end{pmatrix}=\begin{pmatrix}
0 & \mathbb{G}_{0,12}^{} & \cdots & \mathbb{G}_{0,1N}^{}\\
\mathbb{G}_{0,21}^{} & 0 & \ddots & \vdots\\
\vdots & \vdots & \ddots & \mathbb{G}_{0,(N-1)N}^{}\\
\mathbb{G}_{0,N1}^{} & \mathbb{G}_{0,N2}^{} & \cdots & 0
\end{pmatrix}\mathbb{A}^{-1},
\label{double_sources}
\end{equation}
where the matrix $\mathbb{A}$ including many-body interaction is defined as
\begin{equation}
\mathbb{A}=\mathbb{I}_{6N}^{}-\begin{pmatrix}
0 & \boldsymbol{\alpha}_{1}^{}\mathbb{G}_{0,12}^{} & \cdots & \boldsymbol{\alpha}_{1}^{}\mathbb{G}_{0,1N}^{}\\
\boldsymbol{\alpha}_{2}^{}\mathbb{G}_{0,21}^{} & 0 & \ddots & \vdots\\
\vdots & \vdots & \ddots & \boldsymbol{\alpha}_{N-1}^{}\mathbb{G}_{0,(N-1)N}^{}\\
\boldsymbol{\alpha}_{N}^{}\mathbb{G}_{0,N1}^{} & \cdots & \boldsymbol{\alpha}_{N}^{}\mathbb{G}_{0,N(N-1)}^{} & 0
\end{pmatrix},
\label{matrix_interaction}
\end{equation} 
with $\boldsymbol{\alpha}_{i}^{}$ defined as
\begin{equation}
\boldsymbol{\alpha}_{i}^{}=\begin{pmatrix}
\epsilon_0^{}\alpha_{E}^{i}\mathbb{I}_{3}^{} & 0\\
0 & \alpha_{H}^{i}\mathbb{I}_{3}^{}
\end{pmatrix}.
\label{matrix_polarizability}
\end{equation}

To finish, the many-body Green function $G^{\nu\tau}(\mathbf{r},\mathbf{r}_i)\equiv G^{\nu\tau}_{\mathbf{r}\mathbf{r}_i}$ we are looking for to calculate the RTE can also be written in the compact form
\begin{equation}
\mathbb{G}_{\textbf{r}\textbf{r}_i^{}}^{}=\begin{pmatrix}
\mu_{0}^{}\omega^{2}G_{\textbf{r}\textbf{r}_i}^{EE} & \mu_{0}^{}\omega G_{\textbf{r}\textbf{r}_i}^{EM}\\
k\omega G_{\textbf{r}\textbf{r}_i}^{ME} & k^{2}G_{\textbf{r}\textbf{r}_i}^{MM}
\end{pmatrix}.
\label{Compact_system_green_function_single_source}
\end{equation}

The function $\mathbb{G}_{\textbf{r}\textbf{r}_1^{}}^{}$ is the solution of the equation 

\begin{equation}
\left(\mathbb{G}_{\textbf{r}\textbf{r}_1^{}}^{}, \mathbb{G}_{\textbf{r}\textbf{r}_2^{}}^{},\cdots \mathbb{G}_{\textbf{r}\textbf{r}_N^{}}^{}\right)=\left(\mathbb{G}_{0,\textbf{r}\textbf{r}_1^{}}^{}, \mathbb{G}_{0,\textbf{r}\textbf{r}_2^{}}^{},\cdots \mathbb{G}_{0,\textbf{r}\textbf{r}_N^{}}^{}\right)\mathbb{A}^{-1},
\label{single_source}
\end{equation}  
where the matrix $\mathbb{A}$ can be found in Eq.~(\ref{matrix_interaction}) and the source term $\mathbb{G}_{0,\textbf{r}\textbf{r}_i^{}}^{}$ is defined as 
\begin{equation}
\mathbb{G}_{0,\textbf{r}\textbf{r}_i^{}}^{}=\begin{pmatrix}
\mu_{0}^{}\omega^{2}G_{0,\textbf{r}\textbf{r}_i}^{EE} & \mu_{0}^{}\omega G_{0,\textbf{r}\textbf{r}_i}^{EM}\\
k\omega G_{0,\textbf{r}\textbf{r}_i}^{ME} & k^{2}G_{0,\textbf{r}\textbf{r}_i}^{MM}
\end{pmatrix},
\label{Compact_free_space_green_function_single_source}
\end{equation}
where $G_{0,\textbf{r}\textbf{r}_i}^{\mu\nu}\equiv G_{0}^{\mu\nu}(\textbf{r}-\textbf{r}_i)$.



\providecommand{\noopsort}[1]{}\providecommand{\singleletter}[1]{#1}%

\end{document}